\documentclass[pdflatex,sn-mathphys-num]{sn-jnl}

\usepackage{graphicx}%
\usepackage{multirow}%
\usepackage{amsmath,amssymb,amsfonts}%
\usepackage{amsthm}%
\usepackage{mathrsfs,mathtools}%
\usepackage[title]{appendix}%
\usepackage{xcolor}%
\usepackage{textcomp}%
\usepackage{manyfoot}%
\usepackage{booktabs}%
\usepackage{algorithm}%
\usepackage{algorithmicx}%
\usepackage{algpseudocode}%
\usepackage{listings}%

\usepackage{siunitx}
\usepackage{bm}
\usepackage{upgreek}
\usepackage{empheq}
\usepackage{subcaption}
\usepackage{tikz}
\newcommand{\circled}[1]{%
  \tikz[baseline=(char.base)]\node[draw,circle,inner sep=1pt, line width=1pt, solid](char){#1};%
}
\usepackage{comment}
\usepackage{notations}

\theoremstyle{thmstyleone}%
\theoremstyle{thmstyletwo}%

\theoremstyle{thmstylethree}%

\begin{document}

\title[Article Title]{A strategy with reduced models dedicated to parametrized nonlinear strongly coupled thermo-poroelasticity problems}


\author*[1]{\fnm{\'Elise} \sur{Foulatier}}\email{elise.foulatier@ens-paris-saclay.fr}

\author[1]{\fnm{David} \sur{Néron}}\email{david.neron@ens-paris-saclay.fr}
\equalcont{These authors contributed equally to this work.}

\author[1]{\fnm{François} \sur{Louf}}\email{francois.louf@ens-paris-saclay.fr}
\equalcont{These authors contributed equally to this work.}

\author[1]{\fnm{Pierre-Alain} \sur{Boucard}}\email{pierre-alain.boucard@ens-paris-saclay.fr}
\equalcont{These authors contributed equally to this work.}

\affil*[1]{\orgdiv{Université Paris-Saclay, CentraleSupélec, ENS Paris-Saclay, CNRS}, \orgname{LMPS - Laboratoire de Mécanique Paris-Saclay}, \orgaddress{\street{4, avenue des Sciences}, \city{Gif-sur-Yvette}, \postcode{91190}, \country{France}}}


\abstract{This paper offers an approach to deal with parametrized nonlinear strongly coupled thermo-poroelasticity problems. The approach uses the LATIN-PGD method and extends previous work in multiphysics problems. Proper Generalized Decomposition (PGD) allows the building of independent reduced-order bases for each physics. This point is particularly appropriate for thermo-poroelasticity problems whose physics present different dynamics. In parametrized problems dealing with material variability, a new computation is initialized with the result of a previous simulation to speed up the computation times. As a first step, the solver is validated on a standard benchmark in thermo-poroelasticity. The solver shows good performance even in the nonlinear frame. Then, the approach for parametrized problems is addressed on an academic problem and a more complex one, which is part of an industrial process. The results show that the method is effective and less time-consuming than naive approaches.}




\maketitle

\section{Introduction}\label{sec1}

Thermo-poroelasticity problems handle coupling between the three following aspects: solid mechanics, fluid mechanics, and thermics. The modeling of such phenomena is used in sundry domains from biomechanics \cite{andreozzi_2019} to geomechanics \cite{cui_2018,jin_2025} including nuclear waste storage \cite{giot_2018} or reservoirs simulation \cite{wang_2021}. Simulating strongly coupled thermo-poroelasticity leads to the resolution of nonlinear ill-conditioned systems. Consequently there is a need to develop suitable approaches to solve this kind of problems. The first class of methods, called the \textit{monolithic approach} \cite{ordonez_2023,zhang_coupling_2024,mclean_2024}, solves all physics simultaneously at each time step. Although the method provides an accurate solution, the mediocre numerical properties of the system lead to high computational costs. As a result, a second class of methods dedicated to multiphysics problems called the \textit{staggered approach} has been developed in \cite{felippa_1988}. This approach involves the separate resolution of each physics and the exchange of information between physics at each time step. Despite providing smaller and cheaper problems to solve, the method often exhibits an accuracy deficiency. Therefore, the method was improved with an enhanced staggered approach performing interfield iterations at each time step \cite{felippa_2001,vila-cha_2023}. The latter method was adapted to thermo-poroelasticity problems as well \cite{brun_monolithic_2020,kim_2018}.

With the rise of numerical studies, there is an increasing interest in optimization problems. The parameters to optimize typically relate to the material properties, the geometry, or the boundary conditions. A simple approach consists of performing a simple computation per set of parameters. However, such an approach induces high computational costs. Several model order reduction techniques exist to reduce computational costs. The first class of techniques is defined as \textit{a posteriori} methods. Such methods are composed of two phases: an \textit{offline} phase and an \textit{online} phase. The offline phase consists of computing high-fidelity solutions for specific sets of parameters called \textit{snapshots} and using them to compute a \textit{Reduced-Order Basis} (ROB) to approximate the parametric space. During the online phase, the equations are projected on the ROB to compute a fast and accurate approximation of a problem corresponding to a new set of parameters. Different projection based methods are described in \cite{benner_2015}. Among those methods, the \textit{Proper Orthogonalized Decomposition} (POD) \cite{chatterjee_2000} is widely used to compute reduced basis in the time domain. POD has been used in the field of thermo-poroelasticity with the staggered approach \cite{florez_linear_2017}. The critical point with \textit{a posteriori} methods lies in the offline phase. The ROB built during this phase should indeed represent well the parametric space in order to capture the behavior during the online phase. Therefore, it is necessary to choose snapshots in a relevant manner. The \textit{Reduced Basis Method} (RBM) \cite{maday_2006} enables to automatize the choice of those snapshots and thus provides an improvement to the POD methods. Later, the method has been improved by providing an error estimator, which is useful for two main reasons. First, the estimator allows to choose well the snapshots during the \textit{offline} phase. Secondly, the estimator ensures that the solution computed during the \textit{online} phase is accurate enough according to a defined convergence criterion. One can refer to \cite{quarteroni_2011} for more details on the \textit{Certified Reduced Basis Methods}.

The second class of model order reduction techniques is the \textit{a priori} ones. While \textit{a posteriori} methods provide a basis that remains fixed after the offline phase, \textit{a priori} methods offer the possibility to enrich the basis during a computation \cite{ryckelynck_2006}. A method belonging to the \textit{a priori} model order reduction techniques called \textit{Proper Generalized Decomposition} (PGD) has been developed over the last decades \cite{ladeveze_1999,chinesta_2011}. This method relies on the use of the operators in order to build a reduced-order basis. Although the method was initially developed in the frame of nonlinear structural mechanics problems, it has been shown that it is well-adapted to multiphysics problems \cite{beringhier_2010,qin_2015,schuler_2022}. Two possibilities are offered to solve parametrized problems with the PGD. A first approach defines the parameters as an additional coordinate of the problem, which leads to writing the solution in a time-space-parameter decomposition form \cite{pruliere_2010,chinesta_2010}. With this possible approach, an error estimator can be used \cite{chamoin_2017} in order to certify the PGD-based models built for parametrized problems. On the other hand, a second approach omits the parameters from the time-space decomposition, and one considers then a decomposition per simulation. In order to make the method computationally affordable, it is strongly advised using a basis that can be enriched with new modes for each new calculation. In this context, the order in which simulations are run has an impact on performance and strategies may be required for their sequencing.

Such a strategy has been developed based on the iterative LATIN solver \cite{ladeveze_1985}. This iterative solver consists of an initialization providing a time-space solution and iterations correcting the previous solution over the whole time range. The LATIN solver is designed to natively include PGD \cite{ladeveze_1999}, thus giving back a space-time decomposition of the primal fields. The LATIN-PGD solver was used in the domain of multiphysics problems, mainly for linear problems involving two physics, such as poroelasticity \cite{dureisseix_2003} or thermoelasticity \cite{wurtzer_2024}. This strategy has been seldom chosen for nonlinear thermo-poroelasticity \cite{neron_2007}. However, none of these studies develops a many-query framework to solve nonlinear parametrized problems. The computation of parametrized problems using the LATIN method has been studied in \cite{boucard_2003} in the field of contact problems. The main idea introduced consists of reusing the solution of the previous computation as an initialization of the new problem, which leads to a significant improvement in computation times compared to a ``brutal force'' method. Moreover, it is possible to reuse the basis from one computation to another, as shown in \cite{boucard_1999}. The idea is still to use the previous solution as an initial guess but also to take advantage of the previously computed basis and possibly fill it with new modes. The approach proved to be more suitable than POD methods for the case of diffusion problems \cite{heyberger_2012}. Reusing the solution and the basis of previous computations requires ordering the computations in a relevant way. While it is possible to track the parameters space step-by-step \cite{boucard_1999,scanff_2022}, such choice appears irrelevant in specific cases. For instance, another approach was developed in \cite{daby_2025} in order to tackle loading variability. Another way of exploring the parameter space to build a reduced-order basis with a minimal amount of parameters was developed in \cite{heyberger_2013}. When considering material properties, the step-by-step approach remains the most straightforward and powerful method. The use of LATIN-PGD method in the context of many-query studies turns out to be robust for complex problems as shown with nonlinear problems in \cite{neron_2015} or with multiphysics problems \cite{wurtzer_2024}. The multiphysics problems consist in linear strongly coupled thermoelasticity. First, the article highlights the strength of the LATIN-PGD solver against classical direct solvers used in the monolithic approach. Secondly, solving parameterized problems is very powerful when using the LATIN-PGD solver adapted to the many-query context.

This paper extends the work presented in \cite{wurtzer_2024}. Until now, the approach has been applied to linear problems that couple only two physics, and too few studies consider complex multiphysics problems. This paper therefore proposes an approach for solving nonlinear and strongly coupled thermo-poroelasticity problems such as \cite{dureisseix_2003,neron_2007}. The article aims to present a method using model order reduction techniques to reduce the costs of parametrized simulations in thermo-poroelasticity. Moreover, emphasis will be placed on a specific point of the method that strongly affects the convergence speed. Multiphysics problems require determining suitable characteristic times for the expression of search directions. The latter improvements enable us to find analytically an order of magnitude of the characteristic times and avoid technical difficulties linked to using the method in the context of multiphysics problems. 

The paper is organized as follows. In \autoref{sec2}, the thermo-poroelasticity governing equations are presented in a continuous way followed by a discretized way using a finite element scheme. The monolithic solver will also be tackled in this section. \autoref{sec3} focuses on the LATIN-PGD solver. The paper will emphasize some recent advances with the LATIN-PGD solver, providing robustness to the method. \autoref{sec4} presents results on two examples: an academic test case and another representing an industrial application. For both studies, the efficiency of the solver will be validated for a single computation and then in the context of many-query studies. Finally, \autoref{sec5} provides some concluding remarks and outlooks for this work.

\section{Description of the thermo-poroelasticity problem}\label{sec2}

This section focuses on a classical method for dealing with nonlinear strongly coupled thermo-poroelasticity problems. First, we present the governing equations. Secondly, we give the associated finite element formulation. We finally emphasize the monolithic solver. All notations used in this section are summarized in \autoref{tab:notations}.

\subsection{Description of the porous medium}\label{sec21}

A thermo-poroelasticity problem consists of studying a porous medium subjected to thermal loading. A porous medium is modeled as a linear elastic skeleton with pores containing fluid. The porosity of the medium, denoted $\porosity$, quantifies the ratio between the pores volume above the total volume (equal to the pores volume added to the skeleton volume). The medium is considered at the macroscopic scale, and one can thus place itself in the frame of continuum mechanics theory. 

First, we focus on the equivalent material properties of the porous medium. As it is composed of two phases---a fluid one and a solid one---it is necessary to consider properties of both phases to build an accurate homogenized model. In the following, quantities will be expressed depending on different moduli. Under the isotropic hypothesis, the drained bulk modulus is computed with $\drainedbulkmod = E/3(1-2\nu)$, where $E$ is the Young modulus and $\nu$ the Poisson coefficient. The bulk modulus of the fluid phase is denoted $\bulkmodflu$, and the one of the solid phase is denoted $\bulkmodsol$. M.A. Biot introduces in \cite{biot_1941} two quantities to describe a porous medium: the Biot coefficient $\couplP$ and the Biot modulus $\biotmod$. The first one characterizes the coupling between fluid and solid phases and is given by $\couplP = 1- \drainedbulkmod/\bulkmodsol$. The inverse of the second quantity relates linearly the pressure variation to the porosity variation and is defined as $\invbiotmod = \frac{\porosity}{\bulkmodflu} + \frac{\couplP - \porosity}{\bulkmodsol}$. The intrinsic permeability of the medium is denoted $\intpermeability$. The permeability $\permeability$ is commonly used in the fluid equations. This latter is given by: $H = \frac{\intpermeability}{\fluidvisco} = \frac{\hydconductivity}{\densityflu g}$, with $\fluidvisco$ the dynamic viscosity of the fluid and $\hydconductivity$ the hydraulic conductivity. To characterize the thermal behavior, one combines the fluid and solid quantities \cite{coussy_2004}. The heat capacity of the porous medium is given by: $\genheatcapa = \densitysol \heatsol + \porosity \densityflu \heatflu$. The thermal conductivity of the porous medium is defined as: $\thermconductivity = \porosity \thermconductivityflu + (1-\porosity)\thermconductivitysol$. The equivalent thermal expansion coefficient is a combination of both phases' expansion coefficients: $\expcoef = \porosity\expcoefflu + (1-\porosity)\expcoefsol$.

\subsection{Governing equations}\label{sec22}

Having introduced the material properties of the porous medium, one can now focus on the equations governing the problem. The following quantities characterize the nonlinear isotropic homogeneous thermo-poroelasticity problem:
\begin{itemize}
    \item for the solid part: the displacement $\U$ associated with the strain tensor $\Eps$ and the stress tensor $\Sig$;
    \item for the fluid part: the pore pressure $\p$ associated with Darcy's velocity $\vect{V}$;
    \item for the thermal part: the temperature increment $\T = T-\To$ associated with the heat flux $\qth$, where $\To$ is the reference temperature. 
\end{itemize}

Conservation principles govern each of these physics. We consider the momentum balance equation for the solid part, the mass balance equation for the fluid part, and the energy balance equation for the thermal part \cite{lewis_schrefler_1999}. The principles lead to the global admissibility equations detailed for each physics hereinafter. We carry out the study on a bounded domain $\dom \subset \Reel^3$ during the time interval $\I$. In the following, we mention the Sobolev space $\spaceH$, which corresponds to the space of square-integrable functions whose first-order derivatives are also square-integrable. Here, one considers a medium initially at pressure $\po$ and temperature $\To$.

\subsubsection{Solid physics admissibility}\label{sec221}

The mechanical loading can be decomposed into:
\begin{itemize}
    \item a prescribed displacement $\Ud$ on a first part $\bord{u}$ of the domain boundary $\bord{}$;
    \item prescribed external forces $\Fd$ on the complementary part $\bord{F}$ of the domain boundary $\bord{}$;
    \item a prescribed body force $\fd$ in the domain $\dom$.
\end{itemize}

We define the functional space $\mathcal{U}$ as $\mathcal{U} = \{\, \U \mid \U \in \spaceH, \U = \Ud \; \text{on} \: \bord{u} \,\}$ and the associated homogeneous space $\mathcal{U}_0$.

To ensure solid admissibility, one has to respect the kinematic equations \eqref{eq:solid_admissibility_kin}:
\begin{equation}\label{eq:solid_admissibility_kin}
\left\{
\begin{aligned}
    & \U \in \mathcal{U} \\
    & \Eps = \frac{1}{2}(\gradVec{\U} + \gradVec{\U} \Transpose{})\quad\text{in}\: \Ixdom
\end{aligned}
\right.
\end{equation}
and the balance equations \eqref{eq:solid_admissibility_stat}, including the momentum balance equation:
\begin{equation}\label{eq:solid_admissibility_stat}
\left\{
\begin{aligned}
    & \divergTens{\Sig} + \fd = \vect{0} \quad \text{in}\: \Ixdom \\
    & \Sig \normal  = \Fd \quad \text{on}\: \bord{F}\times\I
\end{aligned}
\right.
\end{equation}

\subsubsection{Fluid physics admissibility}\label{sec222}

The fluid loading can be decomposed into:
\begin{itemize}
    \item a prescribed pore pressure $\pd$ on a first part $\bord{p}$ of the domain boundary $\bord{}$;
    \item a prescribed fluid flux $\fluxvd$ on the complementary part $\bord{v}$ of the domain boundary $\bord{}$.
\end{itemize}

We define the functional space $\mathcal{P}$ as $\mathcal{P} = \{\, \p \mid \p \in \spaceH, \p = \pd \; \text{on} \: \bord{p} \,\}$ and the associated homogeneous space $\mathcal{P}_0$.

To ensure fluid admissibility, one has to respect the equations \eqref{eq:fluid_admissibility_kin} related to the primal field:
\begin{equation}\label{eq:fluid_admissibility_kin}
\left\{
\begin{aligned}
    & \p \in \mathcal{P} \\
    & \Z = \grad{\p} \quad \text{in}\: \Ixdom
\end{aligned}
\right.
\end{equation}
and the balance equations \eqref{eq:fluid_admissibility_stat}, including the mass balance equation:
\begin{equation}\label{eq:fluid_admissibility_stat}
\left\{
\begin{aligned}
    & -\divergVec{\vect{V}} = \q \quad\text{in}\: \Ixdom \\
    & \vect{V} \cdot \normal  = \fluxvd \quad\text{on}\: \bord{v}\times\I
\end{aligned}
\right.
\end{equation}

\subsubsection{Thermal physics admissibility}\label{sec223}

The thermal loading can be decomposed into:
\begin{itemize}
    \item a prescribed temperature increment $\Td$ on a first part $\bord{\theta}$ of the domain boundary $\bord{}$;
    \item a prescribed heat flux $\qthd$ on the complementary part $\bord{q}$ of the domain boundary $\bord{}$;
    \item a prescribed heat source $\Rd$ in the domain $\dom$.
\end{itemize}

We define the functional space $\mathcal{T}$ as $\mathcal{T} = \{\, \T \mid \T \in \spaceH, \T = \Td \; \text{on} \: \bord{\theta} \,\}$ and the associated homogeneous space $\mathcal{T}_0$.

To ensure thermal admissibility, one has to respect the equations \eqref{eq:thermics_admissibility_kin} related to the primal field:
\begin{equation}\label{eq:thermics_admissibility_kin}
\left\{
\begin{aligned}
    & \T \in \mathcal{T} \\
    & \X = \grad{\T} \quad\text{in}\: \Ixdom
\end{aligned}
\right.
\end{equation}
and the balance equations \eqref{eq:thermics_admissibility_stat}, including the energy balance equation:
\begin{equation}\label{eq:thermics_admissibility_stat}
\left\{
\begin{aligned}
    & -\divergVec{\qth} + r_d = R \quad\text{in}\: \Ixdom \\
    & \qth \cdot \normal  = \qthd \quad\text{on}\: \bord{q}\times\I
\end{aligned}
\right.
\end{equation}

\subsubsection{Constitutive relations}\label{sec224}

The coupling between the physics, and therefore between the equations previously introduced, is achieved by the constitutive relations.

Modified Hooke's law expresses the stress tensor depending on the strain tensor, the pore pressure, and the temperature increment:
\begin{equation}\label{eq:hooke_law}
    \Sig = \Hooke : \Eps - \couplP \p \Id - \couplT \T \Id
\end{equation}
where $\Hooke$ is the Hooke operator, $\couplP$ the Biot coefficient and $\couplT = 3\expcoefsol \bulkmodsol$ is the coupling coefficient between solid and thermics.

Darcy's law relates Darcy's velocity to the pore pressure gradient:
\begin{equation}\label{eq:darcy_law}
    \vect{V} = -H \Z
\end{equation}

We express the fluid accumulation rate as follows:
\begin{equation}\label{eq:fluid_accumulation_law}
    \q = \invbiotmod \pdot + \couplP \Tr\Epsdot - \couplTP \Tdot
\end{equation}

Fourier's law connects the heat flux to the temperature gradient:
\begin{equation}\label{eq:fourier_law}
    \qth = - k \X
\end{equation}

Finally, the generalized heat source is given by:
\begin{equation}\label{eq:generalized_heat_law}
    R = \genheatcapa \Tdot + \couplT \To \Tr \Epsdot - \couplTP \To \pdot + \densityflu \heatflu \vect{V}\cdot\X - \frac{1}{\permeability}\left(1-3\expcoefflu\To\right) \vect{V}^2
\end{equation}

Equation \eqref{eq:generalized_heat_law} features two nonlinear terms. The first one, employing the scalar product between Darcy's velocity and temperature gradient, refers to the heat transported by the fluid by a convective process, while the second one models viscous dissipation in the fluid. 

In order to build positive operators later on, we introduce the following dual quantities:
\begin{itemize}
    \item For the fluid physics: $\W = -\vect{V}$;
    \item For the thermal physics: $\Y = -\qth$.
\end{itemize}

\subsubsection{Complete equations system}\label{sec225}

Finally, considering all conservative principles and constitutive relations, one can obtain a coupled system of equations.
We begin with the solid part. Body forces $\fd$ and heat sources $\Rd$ are neglected in the following. If Hooke's law \eqref{eq:hooke_law} is injected in the equilibrium equation, one obtains:
\begin{equation}\label{eq:solid_coupled}
    \divergTens{(\Hooke : \Eps)} - \couplP\grad{\p} - \couplT\grad{\T} = \vect{0} \quad\text{in}\: \Ixdom
\end{equation}

For the fluid part, the Darcy's law \eqref{eq:darcy_law} and the fluid accumulation rate \eqref{eq:fluid_accumulation_law} are combined with the mass balance equation \eqref{eq:fluid_admissibility_stat}, which yields:
\begin{equation}\label{eq:fluid_coupled}
    H \lapl{\p} = \invbiotmod \pdot + \couplP \Tr\Epsdot - \couplTP \Tdot \quad\text{in}\: \Ixdom
\end{equation}

Finally, the thermal equation is obtained by gathering Fourier's law \eqref{eq:fourier_law}, the generalized heat source \eqref{eq:generalized_heat_law} and the energy balance equation \eqref{eq:thermics_admissibility_stat}:
\begin{equation}\label{eq:thermics_coupled}
    k \lapl{\T} = \rho c \Tdot + \couplT \To \Tr \Epsdot - \couplTP\To  \pdot - \rho_F c_F \W\cdot\X - \frac{1}{H}\left(1-3\alpha_F\To \right) \W^2 \quad\text{in}\: \Ixdom
\end{equation}

The strong coupling between all physics is visible in the last equations. All primal fields appear in each of the equations (\ref{eq:solid_coupled}, \ref{eq:fluid_coupled}, \ref{eq:thermics_coupled}), which signifies that all physics are interacting together.


We write the complete problem in its strong formulation hereinafter:

\bigskip
Find $\U \in \espaceU$, $\p \in \espaceP$, $\T \in \espaceT$ such that: 
\begin{subequations}\label{eq:strong_formulation_TPM}
\begin{equation}\label{eq:solid_final}
    \divergTens{(\Hooke : \Eps)} - \couplP\grad{\p} - \couplT\grad{\T} = \vect{0}  \quad\text{in}\: \Ixdom
\end{equation}
\begin{equation}\label{eq:fluid_final}
    -\permeability \lapl{\p} + \invbiotmod \pdot + \couplP \Tr\Epsdot - \couplTP \Tdot = 0  \quad\text{in}\: \Ixdom
\end{equation}
\begin{equation}\label{eq:thermics_final}
- \thermconductivity \lapl{\T} + \genheatcapa \Tdot + \couplT \To \Tr \Epsdot - \couplTP\To  \pdot - \densityflu \heatflu \W\cdot\X - \frac{1}{\permeability}\left(1-3\expcoefflu\To \right) \W^2 = 0  \quad\text{in}\: \Ixdom
\end{equation}
\begin{equation}\label{eq:po_final}
    \p(\M, t=0) = \po \quad \forall \M \in \dom 
\end{equation}
\begin{equation}\label{eq:To_final}
    \T(\M, t=0) = 0 \quad \forall \M \in \dom 
\end{equation}
\end{subequations}

\subsection{Variational formulation} \label{sec23}
The equations given in (\ref{eq:solid_admissibility_stat}, \ref{eq:fluid_admissibility_stat}, \ref{eq:thermics_admissibility_stat}) can be written in a weak form using a variational formulation. Therefore, we rewrite the coupled problem in the subsequent form:

\bigskip
Find $\U \in \espaceU$, $\p \in \espaceP$, $\T \in \espaceT$ such that $\forall t \in \I$: 
\begin{subequations}\label{eq:variational_formulation}
\begin{equation}\label{eq:var_solid}
\forall \Utest \in \espaceUo, \displaystyle\intO{\Sig : \Eps(\Utest)} = \int_{\bord{F}} \scalprod{\Fd}{\Utest} \dS
\end{equation}
\begin{equation}\label{eq:var_fluid}
\forall \Ptest \in \espacePo, \displaystyle\intO{\left(\q \Ptest + \W\cdot\grad{\Ptest}\right)}  = -\int_{\bord{v}} \fluxvd \Ptest \dS
\end{equation}
\begin{equation}\label{eq:var_thermics}
\forall \Ttest \in \espaceTo, \displaystyle\intO{\left(R\Ttest + \Y\cdot\grad{\Ttest} \right)} = -\int_{\bord{q}} \qthd \Ttest \dS
\end{equation}
\begin{equation}\label{eq:var_po}
    \p(\M, t=0) = \po \quad \forall \M \in \dom 
\end{equation}
\begin{equation}\label{eq:var_To}
    \T(\M, t=0) = 0 \quad \forall \M \in \dom 
\end{equation}
\end{subequations}

\subsection{Monolithic solver}\label{sec24}

This section aims to describe the monolithic solver whose solution serves as a reference solution in the article. Dealing with nonlinearities induces low computational efforts with the LATIN-PGD solver compared to the monolithic one. Consequently, we compare the solvers in the frame of linear problems, and we implement the nonlinearities only in the LATIN-PGD solver.

Assuming the problem is linear, one can eliminate the quadratic terms in \eqref{eq:thermics_final}. Hence, the variational formulation aforementioned becomes:

\bigskip
Find $\U \in \espaceU$, $\p \in \espaceP$, $\T \in \espaceT$ such that $\forall t \in \I$: 
\begin{subequations}\label{eq:variational_formulation_linear}
\begin{equation}\label{eq:var_solid_linear}
\forall \Utest \in \espaceUo, \displaystyle\intO{\Hooke : \Eps(\U) : \Eps(\Utest)} - \intO{\couplP \p \Tr\Eps(\Utest)} - \intO{\couplT \T \Tr\Eps{\Utest}} = \int_{\bord{F}} \scalprod{\Fd}{\Utest} \dS
\end{equation}
\begin{equation}\label{eq:var_fluid_linear}
\forall \Ptest \in \espacePo, \displaystyle\intO{\couplP \Tr\Epsdot(\U) \Ptest}+ \intO{\invbiotmod \pdot \Ptest} - \intO{\couplTP\Tdot \Ptest} + \intO{\permeability\grad{\p}\cdot\grad{\Ptest}}  = -\int_{\bord{v}} \fluxvd \Ptest \dS
\end{equation}
\begin{equation}\label{eq:var_thermics_linear}
\forall \Ttest \in \espaceTo, \displaystyle\intO{\couplT \To\Tr\Epsdot(\U) \Ttest} - \intO{\couplTP\To\pdot \Ttest}+ \intO{\genheatcapa \Tdot \Ttest}+ \intO{\thermconductivity\grad{\T}\cdot\grad{\Ttest}} = -\int_{\bord{q}} \qthd \Ttest \dS
\end{equation}
\begin{equation}\label{eq:var_po_linear}
    \p(\M, t=0) = \po \quad \forall \M \in \dom 
\end{equation}
\begin{equation}\label{eq:var_To_linear}
    \T(\M, t=0) = 0 \quad \forall \M \in \dom 
\end{equation}
\end{subequations}

We discretize the thermo-poroelasticity problem in space using the finite element method. Considering the primal fields $\U$, $\p$ and $\T$ as the unknowns and relying on the variational formulation \eqref{eq:variational_formulation_linear}, we write the coupled equations in the matrix form given in \eqref{eq:matrix_system_mono}.
\begin{equation}\label{eq:matrix_system_mono}
    \begin{bmatrix}
        \matD{0} &\matD{0}&\matD{0}\\
        \Mpu{C}&\Mpp{C}&-\Mpt{C}\\
        \Mtu{C} & \Mtp{C} &\Mtt{C}
    \end{bmatrix}\begin{bmatrix}
        \uVdot\\
        \pVdot\\
        \tVdot
    \end{bmatrix} + \begin{bmatrix}
        \Muu{K} & - \Mup{K} & -\Mut{K}\\
        \matD{0}&\Mpp{K}&\matD{0}\\
        \matD{0}&\matD{0}&\Mtt{K}
    \end{bmatrix}\begin{bmatrix}
        \uV\\
        \pV\\
        \tV
    \end{bmatrix}=\begin{bmatrix}
        \fu{f}\\
        \fp{f}\\
        \ft{f}
    \end{bmatrix}
\end{equation}

Matrices $\Muu{K}$, $\Mpp{K}$, and $\Mtt{K}$ refer to the stiffness matrices corresponding to the solid, the fluid, and the thermal physics. $\Mpp{C}$ and $\Mtt{C}$ designate the conductivity matrix for the fluid and thermal parts. All other matrices represent the coupling between the different physics. Finally, the right-hand side vector is composed of the external loadings related to the three physics. The detailed expressions of those matrices are specified in \cite{nonino_2022}.

The matrix form of the problem given in \eqref{eq:matrix_system_mono} corresponds to a semi-discretized form as we have only discretized $\dom$. We have recourse to a backward Euler scheme for the temporal integration.

The monolithic solver is designed to solve the three physics simultaneously, as done in \cite{ordonez_2023}. Such an approach leads to accurate results but meanwhile features a lack of modularity as mentioned in \cite{felippa_2001}. This is due to the coincidental presence of all physics, which makes it challenging to add slight changes to one physics (adding nonlinear effects, for instance). In parametrized problems, one aims to build a reduced basis approximating each physics well. The monolithic solver enables the building of three reduced bases with \textit{a posteriori} model order reduction technique. However, the bases will all contain the same number of modes, while other algorithms present the advantage of building a basis independently and, therefore, reducing the number of modes. Thus, we present hereinafter a solver offering greater modularity.
\section{The LATIN-PGD solver for nonlinear thermo-poroelasticity}\label{sec3}

This section presents the LATIN-PGD approach for strongly coupled nonlinear thermo-poro-\linebreak[1]\hspace{0pt}elasticity problems. The LATIN-PGD method is a non-incremental solver initially developed in the frame of nonlinear mechanics in \cite{ladeveze_1999}. It has been subsequently extended to several case studies, including multiphysics problems, firstly presented in \cite{dureisseix_2003} for poroelasticity problems. The LATIN-PGD method offers two major variants: the functional formulation and the internal variable formulation. It has been shown in \cite{scanff_2021} that the internal variables formulation is more expensive regarding the computation times and the memory. Thus, we would rather use the functional formulation than the internal variables. The LATIN-PGD method in its functional formulation applied to multiphysics problems is presented in \cite{wurtzer_2024} for the specific case of strongly coupled thermoelastic problems. We now introduce the solver for the case of nonlinear strongly coupled thermo-poroelasticity problems.

\subsection{Solver description}\label{sec31}

The LATIN-PGD method relies on the separation of difficulties. As explained in \cite{dureisseix_2003}, the equations are divided into two sets. The first set of equations denoted $\Ad$, gathers the decoupled and linear equations, possibly global. This includes the admissibility equations (\ref{eq:solid_admissibility_kin}, \ref{eq:solid_admissibility_stat}, \ref{eq:fluid_admissibility_kin}, \ref{eq:fluid_admissibility_stat}, \ref{eq:thermics_admissibility_kin}, \ref{eq:thermics_admissibility_stat}). The second set of equations, denoted $\Gam$, is composed of the local equations, possibly coupled and nonlinear. This includes here the constitutive equations (\ref{eq:hooke_law}, \ref{eq:darcy_law}, \ref{eq:fluid_accumulation_law}, \ref{eq:fourier_law}, \ref{eq:generalized_heat_law}). These latter equations are time-dependent, which implies that initial conditions \eqref{eq:po_final} and \eqref{eq:To_final} are also contained in $\Gam$. The solution $\sref$ of the multiphysics problem corresponds then to the intersection of the two sets of equations:
\begin{equation}
    \sref = \Ad \cap \Gam
\end{equation}

The algorithm iteratively searches solutions in $\Ad$ and $\Gam$. The quantities belonging to $\Gam$ are denoted $\locShort{\square}$. Each iteration $n+1$ is decomposed in two stages:
\begin{itemize}
    \item a \textbf{coupled stage} enabling to get a solution $\loc{\s} \in \Gam$ from a known solution $\linn{\s} \in \Ad$ using a search direction $\Eup$;
    \item a \textbf{decoupled stage} enabling to get a solution $\lin{\s} \in \Ad$ from a known solution $\loc{\s} \in \Gam$ using a search direction $\Edo$.
\end{itemize}
The search directions $\Eup$ and $\Edo$ are given by:
\begin{subequations}
\begin{equation}
\Eup \equiv \left\{
\Delta \s = (\Delta \sM,\Delta \sF, \Delta \sT) \quad \left|
\begin{array}{lllll}
    \Delta \Sig + \Hes^+ \mathbin{:} \Delta\Eps &= \tens{0} \\
    \Delta \W + \Hzw^+ \Delta\Z &= \vect{0} \\
    \Delta \q + \Hpq^+ \Delta\p &= 0 \\
    \Delta \Y + \Hxy^+ \Delta\X &= \vect{0} \\
    \Delta \R + \Htr^+ \Delta\T &= 0
\end{array}\right.
\right\}
\end{equation}
\begin{equation}
\Edo \equiv \left\{ \Delta \s = (\Delta \sM,\Delta \sF, \Delta \sT) \quad \left|
\begin{array}{lllll}
    \Delta \Sig - \Hes^- \mathbin{:}\Delta\Eps &= \tens{0} \\
    \Delta \W - \Hzw^- \Delta\Z &= \vect{0} \\
    \Delta \q - \Hpq^- \Delta\p &= 0 \\
    \Delta \Y - \Hxy^- \Delta\X &= \vect{0} \\
    \Delta \R - \Htr^- \Delta\T &= 0
\end{array}\right. 
\right\}
\end{equation}   
\end{subequations}
where  $\Delta\square$ designates the variation of the quantity $\square$ between two consecutive stages. The aforementioned equations are written in a general frame. However, the search directions are key parameters of the method, and their choice greatly influences the convergence speed of the algorithm. Based on the constitutive equations (\ref{eq:hooke_law}, \ref{eq:darcy_law}, \ref{eq:fluid_accumulation_law}, \ref{eq:fourier_law}, \ref{eq:generalized_heat_law}), the parameters can be identified as follows:
\begin{align}
    \label{eq:Hes} &\Hes^+ = \Hes^- = \Hes = \Hooke\\
    \label{eq:Hzw} &\Hzw^+ = \Hzw^- = \Hzw = \permeability \Id\\
    \label{eq:Hpq} &\Hpq^+ = \Hpq^- = \Hpq = \invtf\invbiotmod\Id\\
    \label{eq:Hxy} &\Hxy^+ = \Hxy^- = \Hxy = \thermconductivity \Id\\
    \label{eq:Htr} &\Htr^+ = \Htr^- = \Htr = \invttheta \genheatcapa \Id
\end{align}
The choice of the characteristic times $\tf$ and $\ttheta$ will be clarified in \autoref{sec36}.

We now detail the equations underlying the algorithm in the following sections.

\subsection{Coupled stage}\label{sec32}

Knowing the solution from the previous iteration $\linn{\s} \in \Ad$, one computes the solution $\loc{\s} \in \Gam$. The search direction $\Eup$ yields:
\begin{subequations}
\begin{empheq}[left=\empheqlbrace]{align}
    &\loc{\Sig} + \Hes \mathbin{:}\loc{\Eps} = \underbrace{\linn{\Sig} + \Hes\mathbin{:}\linn{\Eps}}_{\displaystyle\linn{\Atens}} \\
    &\loc{\W} + \Hzw \loc{\Z} =  \underbrace{\linn{\W} + \Hzw\linn{\Z}}_{\displaystyle\linn{\betavec}} \\
    &\loc{\q} + \Hpq \loc{\p} = \underbrace{\linn{\q} + \Hpq\linn{\p}}_{\displaystyle\linn{\gammascal}} \\
    &\loc{\Y} + \Hxy \loc{\X} = \underbrace{\linn{\Y} + \Hxy \linn{\X}}_{\displaystyle\linn{\deltavec}} \\
    &\loc{\R} + \Htr \loc{\T} = \underbrace{\linn{\R} + \Htr\linn{\T}}_{\displaystyle\linn{\phiscal}}  
\end{empheq}  
\end{subequations}

Using the constitutive equations, one gets:
\begin{subequations}
\begin{empheq}[left=\empheqlbrace]{align}
    \label{eq:coupled1} &(\Hooke + \Hes)\mathbin{:}\loc{\Eps} - \couplP\loc{\p}\Id - \couplT\loc{\T}\Id =  \linn{\Atens} \\
    \label{eq:coupled2} &(\bm{\permeability} + \Hzw)\loc{\Z} = \linn{\betavec} \\
    \label{eq:coupled3} &\invbiotmod\locdot{\p} + \couplP \Tr{\locdot{\Eps}} - \couplTP \locdot{\T} + \Hpq \loc{\p} = \linn{\gammascal} \\
    \label{eq:coupled4} &\Big(\thermconductivity + \Hxy\Big)\loc{\X} = \linn{\deltavec} \\
    \label{eq:coupled5} &\begin{aligned}[t]
    \genheatcapa \locdot{\T} + \couplT \To \Tr{\locdot{\Eps}} - \couplTP \To \locdot{\p} &- \densityflu \heatflu \To \loc{\W} \cdot \loc{\X} \\
    &- \frac{1}{\permeability} \left(1 - 3 \expcoefflu \To \right) \loc{\W}^2 + \Htr \loc{\T} = \linn{\phiscal}
    \end{aligned}
\end{empheq}
\end{subequations}

The gradient of pressure $\loc{\Z}$ and the gradient of temperature $\loc{\X}$ are directly computed from equations \eqref{eq:coupled2} and \eqref{eq:coupled4}. Thus, the two nonlinear terms in \eqref{eq:coupled5} are known quantities that can be transferred to the right-hand side of \eqref{eq:coupled5}. One then computes the remaining quantities by solving the following coupled system of ordinary differential equations \eqref{eq:coupled_ODE}:
\begin{equation}\label{eq:coupled_ODE}
\left\{\begin{aligned}
    &(\Hooke+\Hes)\mathbin{:}\loc{\Eps} - \couplP\loc{\p}\Id - \couplT \loc{\T}\Id = \linn{\Atens} \\
    &\couplP \Tr{\locdot{\Eps}} + \invbiotmod \locdot{p} - \couplTP \locdot{\T}+ \Hpq \loc{\p}  = \linn{\gammascal} \\
    & \begin{aligned}[t]
    \genheatcapa \locdot{\T}+ \couplT \To \Tr{\locdot{\Eps}} - \couplTP\To  \locdot{\p} + \Htr \loc{\T} = \linn{\phiscal} &+\densityflu\heatflu \To \loc{\W}\cdot\loc{\X} \\
    &+\frac{1}{\permeability}\left(1-3\expcoefflu\To \right) \loc{\W}^2
    \end{aligned}
    \end{aligned}\right.
\end{equation}

Finally, one deduces the dual quantities with the search direction $\Eup$:
\begin{equation}
    \left\{\begin{aligned}
        &\loc{\Sig} = \linn{\Atens} - \Hes\mathbin{:}\loc{\Eps}\\
        &\loc{\W} = \linn{\betavec} - \Hzw\loc{\Z}\\
        &\loc{\q} = \linn{\gammascal} - \Hpq\loc{\p}\\
        &\loc{\Y} = \linn{\deltavec} - \Hxy\loc{\X}\\
        &\loc{\R} = \linn{\phiscal} - \Htr\loc{\T}
    \end{aligned}
    \right.
\end{equation}

\subsection{Decoupled stage}\label{sec33}

Knowing $\loc{\s} \in \Gam$ from the previous stage, one looks for $\lin{\s} \in \Ad$. The search direction $\Edo$ yields:
\begin{subequations}
\begin{empheq}[left=\empheqlbrace]{align}
    (&\lin{\Sig} - \loc{\Sig}) - \Hes\mathbin{:}(\lin{\Eps} - \loc{\Eps}) = 0 \\
    (&\lin{\W} - \loc{\W}) - \Hzw(\lin{\Z} - \loc{\Z}) = 0 \\
    (&\lin{\q} - \loc{\q}) - \Hpq(\lin{\p} - \loc{\p}) = 0 \\
    (&\lin{\Y} - \loc{\Y}) - \Hzw(\lin{\X} - \loc{\X}) = 0 \\
    (&\lin{\R} - \loc{\R}) - \Htr(\lin{\T} - \loc{\T}) = 0
\end{empheq}  
\end{subequations}
Therefore, the dual quantities at the decoupled stage are expressed as:
\begin{subequations}
\begin{empheq}[left=\empheqlbrace]{align}
    \label{eq:decoupled1} & \lin{\Sig} = (\underbrace{\loc{\Sig} - \Hes\mathbin{:}\loc{\Eps}}_{\displaystyle\loc{\Atens}}) + \Hes\mathbin{:}\lin{\Eps}\\
    \label{eq:decoupled2} & \lin{\W} = \underbrace{(\loc{\W} - \Hzw\loc{\Z})}_{\displaystyle\loc{\betavec}} + \Hzw\lin{\Z} \\
    \label{eq:decoupled3} & \lin{\q} = \underbrace{(\loc{\q} - \Hpq\loc{\p})}_{\displaystyle\loc{\gammascal}} + \Hpq\lin{\p} \\
    \label{eq:decoupled4} & \lin{\Y} = \underbrace{(\loc{\Y} - \Hxy\loc{\X})}_{\displaystyle\loc{\deltavec}} + \Hxy\lin{\X} \\
    \label{eq:decoupled5} & \lin{\R} = \underbrace{(\loc{\R} - \Htr\loc{\T})}_{\displaystyle\loc{\phiscal}} + \Htr\lin{\T}
\end{empheq}  
\end{subequations}

Let first focus on the \textbf{solid quantities}. When introducing the expression of the stress tensor \eqref{eq:decoupled1} in the solid variational formulation \eqref{eq:var_solid}, the following weak form is obtained:\\
Find $\lin{\U} \in \espaceU$ such that $\forall t \in \I$: 
\begin{equation*}
    \forall\: \Utest \in \espaceUo, \intO{\Eps(\lin{\U})\mathbin{:}\Hes\mathbin{:}\Eps(\Utest)} = \intS{F}{\scalprod{\Fd}{\Utest}} - \intO{\loc{\Atens}:\Eps(\Utest)}
\end{equation*}

At the initialization of the algorithm, one takes $\Atens = \tens{0}$ and the problem is solved in the discrete form \eqref{eq:solid_FE_init}:
\begin{equation}\label{eq:solid_FE_init}
\left\{
    \begin{aligned}
        \Muu{H} \uVO(t) &= \fu{f}(t) \quad \text{in}\:\dom\\
        \uVO(t) &= \uVd(t) \quad \text{on}\: \bord{\U}
    \end{aligned}
\right.
\end{equation}
The matrix $\Muu{H}$ is bound to the search direction $\Hes$. Given the choice of search direction made in \eqref{eq:Hes}, we consider here $\Muu{H} = \Muu{K}$.

For all following LATIN iterations, the correction $\linCorU = \lin{\uV} - \linShort{\uV}$ is computed by solving the following linear system \eqref{eq:decoupled_solid}:
\begin{equation}\label{eq:decoupled_solid}
\left\{
    \begin{aligned}
        \Muu{H} \linCorU(t) &= \fu{g}(t) \quad \text{in}\:\dom\\
        \linCorU(t) &= \vecD{0} \quad \text{on}\: \bord{u}
    \end{aligned}
\right.
\end{equation}
with $\fu{g} = \FAc - \Muu{H}(\linShort{\uV}-\uVO)$ where $\FAc$ refers to the right-hand side computed with the local stage quantities.

Let now consider the \textbf{fluid quantities}. The expressions of $\lin{\W}$ and $\lin{\q}$ given in \eqref{eq:decoupled2} and \eqref{eq:decoupled3} are injected in the fluid variational formulation \eqref{eq:var_fluid} leading to the following problem:\\
Find $\lin{\p} \in \espaceP$ such that $\forall t \in \I$: 
\begin{equation}
\begin{split}
    \forall\: \Ptest \in \espacePo, \intO{\grad{\lin{\p}}\cdot\Hzw\grad{\Ptest}}+&\intO{\lin{p}\Hpq\Ptest}=-\intO{\loc{\betavec}\grad{\Ptest}}\\
        &-\intO{\loc{\gammascal}\Ptest}-\int_{\bord{v}} \fluxvd \Ptest \dS
\end{split}
\end{equation}
As described for the solid quantities, the problem is solved in a discrete form using finite elements. At the initialization of the algorithm, one solves:
\begin{equation}\label{eq:fluid_FE_init}
\left\{
    \begin{aligned}
        \Mpp{H} \pVO(t) &= \fp{f}(t) \quad \text{in}\:\dom\\
        \pVO(t) &= \pVd(t) \quad \text{on}\: \bord{\p}
    \end{aligned}
\right.
\end{equation}
with $\Mpp{H} = \invtf \Mpp{C} + \Mpp{K}$.

For all following LATIN iterations, the correction $\linCorP = \lin{\pV}-\linShort{\pV}$ is computed by solving the linear system \eqref{eq:decoupled_fluid}:
\begin{equation}\label{eq:decoupled_fluid}
\left\{
    \begin{aligned}
        \Mpp{H} \linCorP(t) &= \fp{g}(t) \quad \text{in}\:\dom\\
        \linCorP(t) &= \vecD{0} \quad \text{on}\: \bord{\p}
    \end{aligned}
\right.
\end{equation}
with $\fp{g} = \Fbc + \Fgc - \Mpp{H}(\linShort{\pV}-\pVO)$ where $\Fbc$ and $\Fgc$ refer to the right-hand-side computed with the local stage quantities.

Finally, the \textbf{thermal quantities} are studied. The reasoning is similar to the one used for fluid quantities. One combines the expressions of $\lin{\Y}$ and $\lin{\R}$ with the variational formulation \eqref{eq:var_thermics} and we reduce at the following problem:\\
Find $\lin{\T} \in \espaceT$ such that $\forall t \in \I$: 
\begin{equation}
    \begin{split}
        \forall\: t \in \I, \forall\: \Ttest \in \espaceTo, \intO{\grad{\lin{\T}}\cdot\Hxy\grad{\Ttest}} +& \intO{\lin{\T}\Htr\Ttest} = -\intO{\loc{\deltavec}\grad{\Ttest}}\\
        &-\intO{\loc{\phiscal}\Ttest}-\int_{\bord{q}} \qthd \Ttest \dS
    \end{split}
\end{equation}

The initial field of temperature is computed from the discrete problem \eqref{eq:thermics_FE_init}:
\begin{equation}\label{eq:thermics_FE_init}
\left\{
    \begin{aligned}
        \Mpp{H} \tVO(t) &= \ft{f}(t) \quad \text{in}\:\dom\\
        \tVO(t) &= \tVd(t) \quad \text{on}\: \bord{\T}
    \end{aligned}
\right.
\end{equation}
with $\Mtt{H} = \invttheta \Mtt{C} + \Mtt{K}$.

The correction $\linCorT = \lin{\tV}-\linShort{\tV}$ is computed for following LATIN iterations with the linear system \eqref{eq:decoupled_thermics}:
\begin{equation}\label{eq:decoupled_thermics}
\left\{
    \begin{aligned}
        \Mtt{H} \linCorT(t) &= \ft{g}(t) \quad \text{in}\:\dom\\
        \linCorT(t) &= \vecD{0} \quad \text{on}\: \bord{\T} 
    \end{aligned}
\right.
\end{equation}
with $\ft{g} = \Fdc + \Ffc - \Mtt{H}(\linShort{\tV}-\tVO)$ where $\Fdc$ and $\Ffc$ refer to the right-hand-side computed with the local stage quantities.

\subsection{Stopping criterion}\label{sec34}

We gave the content of an algorithm iteration in \autoref{sec32} for the coupled stage and in \autoref{sec33} for the decoupled stage. One needs now to know when to stop the iterations. For that, we define convergence indicators for each physics, and for each of these indicators, we introduce a norm specific to the physics. 

The \textbf{solid} convergence indicator is given by:
\begin{equation}\label{eq:ind_solid}
    {\etaM} ^2 = \frac{\norme{\lin{\sM}-\loc{\s}}{\Hes}^2}{\frac{1}{2}(\norme{\lin{\sM}}{\Hes}^2 + \norme{\loc{\s}}{\Hes}^2)} \quad \text{with} \norme{\sM}{\Hes}^2 = \intTO{\Eps : \Hes : \Eps}
\end{equation}
The \textbf{fluid} convergence indicator is given by:
\begin{equation}\label{eq:ind_fluid}
    {\etaF} ^2 = \frac{\norme{\lin{\sF}-\loc{\s}}{\Hzw}^2}{\frac{1}{2}(\norme{\lin{\sF}}{\Hzw}^2 + \norme{\loc{\s}}{\Hzw}^2)} \quad \text{with} \norme{\sF}{\Hzw}^2 = \intTO{\Z \cdot \Hzw \Z}
\end{equation}
The \textbf{thermal} convergence indicator is given by:
\begin{equation}\label{eq:ind_thermics}
    {\etaT} ^2 = \frac{\norme{\lin{\sT}-\loc{\s}}{\Hxy}^2}{\frac{1}{2}(\norme{\lin{\sT}}{\Hxy}^2 + \norme{\loc{\s}}{\Hxy}^2)} \quad \text{with} \norme{\sT}{\Hxy}^2 = \intTO{\X \cdot \Hxy \X}
\end{equation}

Equations \eqref{eq:ind_solid}, \eqref{eq:ind_fluid} and \eqref{eq:ind_thermics} show that the indicator quantifies the distance between two successive solutions $\loc{\s}$ and $\lin{\s}$. When this latter is considered sufficiently small, the stopping criterion claims that the approximation of the reference solution is good enough. The user defines a tolerance threshold denoted $\eta_c$ and the algorithm stops when $\eta = \text{max}\left(\etaM, \etaF, \etaT \right) < \eta_c$. If one or more physics converge faster than the others, it stops when its indicator is lower than the tolerance threshold.

\subsection{Proper Generalized Decomposition within the LATIN framework}\label{sec35}
The so-called LATIN-PGD method includes a model order reduction technique at the decoupled stage. This method is the \textit{Proper Generalized Decomposition}---denoted PGD in the following---and belongs to the \textit{a priori} model order reduction techniques. The distinctive feature of the method is that the reduced-order basis is built over the algorithm, which is why one evokes an \textit{on-the-fly} model reduction technique.  When PGD is incorporated into the LATIN solver, one writes the primal fields as a combination of time and space functions. After $n$ iterations of the algorithm, the displacement $\uV$, the pressure $\pV$ and the temperature $\tV$ are written as follows:
\begin{equation}\label{eq:primal_fields_PGD}
    \left\{
    \begin{aligned}
    \linn{\uV} (t) &= \uVO(t) + \sumPGD{i}{k} \TimeModeU{i}{}(t) \SpaceModeU{i} \quad k\leqslant n\\
    \linn{\pV} (t) &= \pVO(t) + \sumPGD{i}{l} \TimeModeP{i}(t) \SpaceModeP{i} \quad l\leqslant n\\
    \linn{\tV} (t) &= \tVO(t) + \sumPGD{i}{m} \TimeModeT{i}(t) \SpaceModeT{i} \quad m\leqslant n
    \end{aligned}
    \right.
\end{equation} 

At each iteration and for each physics, one seeks a new pair of modes in the form given in \eqref{eq:gen_modes_PGD} if needed.
\begin{equation}\label{eq:gen_modes_PGD}
    \left\{
    \begin{aligned}
    \linCorU (t) &= \TimeModeU{}{}(t) \SpaceModeU{}\\
    \linCorP (t) &= \TimeModeP{}(t) \SpaceModeP{}\\
    \linCorT (t) &= \TimeModeT{}(t) \SpaceModeT{}
    \end{aligned}
    \right.
\end{equation}

The decoupled stage begins with an update step that modifies all time functions to improve the solution. The operation is less expensive than generating a new spatial mode. Thus, if the update step adequately improves the solution, no mode is generated afterward.

More details on the PGD when dealing with multiphysics problems solved with a LATIN-PGD solver can be found in \cite{wurtzer_2024}. The equations were presented for the displacement field $\uV$, but it is easily reproducible for the other primal fields $\pV$ and $\tV$.

\subsection{Choice of appropriate search directions}\label{sec36}

We have evoked the choice of search directions in \autoref{sec31}. As a recall, the expressions of the five components of the search directions are given in (\ref{eq:Hes}, \ref{eq:Hzw}, \ref{eq:Hpq}, \ref{eq:Hxy}, \ref{eq:Htr}). One can notice that the expressions for $\Hpq$ and $\Htr$ involve two characteristic times $\tf$ and $\ttheta$, which need to be determined to optimize the convergence speed of the algorithm. We deepen the reasoning proposed in \cite{neron_2008} to address more complex problems with strong coupling and complex geometries. The key point of the method relies on an approximation of the coupled system of equations (\ref{eq:solid_final}, \ref{eq:fluid_final}, \ref{eq:thermics_final}) given in \autoref{sec225}. For the sake of simplicity, we will first present the method on a simple monodimensional poroelasticity problem. As only two physics are coupled, the problem involves one characteristic time. After validating the method on the monodimensional problem, we suggest an approach to tackle strongly coupled nonlinear thermo-poroelasticity problems implicating two characteristic times.

\subsubsection{Application to a monodimensional poroelasticity problem}\label{sec361}

We consider here the poroelasticity example proposed in \cite{neron_2008}. As a recall, one considers a beam of cross-section $S$ clamped at one end and submitted to a mechanical loading $F(t)$ at the other. The fluid pressure is imposed at both ends of the beam. We remove all terms related to thermal physics to obtain the poroelasticity equations. Moreover, we simplify the equations to consider a monodimensional example. Therefore, we obtain the following system of equations, modeling a coupled poroelasticity problem:
\begin{equation}\label{eq:1D_poroelasticity_continuous}
    \left\{
    \begin{aligned}
    &E\varepsilon - \couplP \p = F(t)/S \\
    &\couplP \Dot{\varepsilon} + \invbiotmod \pdot - \permeability \derivPart{^2 \p}{x^2} = 0
    \end{aligned}
    \right.
\end{equation}
We differentiate the first equation and neglect the temporal derivative of $F(t)$. Thus, when combining both equations, one gets a parabolic partial derivative equation of the form:
\begin{equation}\label{eq:parabolic_pde_pressure}
    \left(\frac{\couplP^2}{E} +\invbiotmod\right) \pdot - \permeability \derivPart{^2 \p}{x^2} = 0
\end{equation}
The solution is sought as a product of space and time functions: $\p(x,t) = f(x)g(t)$. One can show by solving two ordinary differential equations that $f(x) = f\left(\frac{x}{L_c}\right)$ is an oscillating function where $L_c$ designates a characteristic length and that $g(t) = g\left(\frac{t}{\tf}\right)$ is an exponential form where $\tf$ designates a characteristic time.

We express the strain as $\Dot{\varepsilon} = \frac{\couplP}{E}\frac{\p}{\Delta t}$ where $\Delta t$ designates the size of a time step.
Moreover, we inject the form of the pore pressure in the second equation of \eqref{eq:1D_poroelasticity_continuous}:
\begin{equation}\label{eq:1D_poroelasticity_approx}
    \left\{
    \begin{aligned}
    &\Dot{\varepsilon} = \frac{\couplP}{E}\frac{\p}{\Delta t} \\
    &\couplP \Dot{\varepsilon} - \invbiotmod \invtf \p - \permeability \left(-\frac{1}{L_c^2}\right)\p = 0
    \end{aligned}
    \right.
\end{equation}

Finally, we get the following expression of $\tf$:
\begin{equation}\label{eq:1D_poroelasticity_tf}
    \tf = \invbiotmod \frac{1}{\frac{\permeability}{L_c^2} + \frac{\couplP^2}{E}\frac{1}{\Delta t}}
\end{equation}
As a recall, the analytical value presented in \cite{neron_2008} is given by:
\begin{equation}\label{eq:tf_DN_2008}
    \tf = \invbiotmod \frac{L_c^2}{\permeability}
\end{equation}
The analytical value of $\tf$ given in \cite{neron_2008} thus neglects the coupling effects, as one recovers a simplification of the expression \eqref{eq:1D_poroelasticity_tf} without the term $\frac{\couplP^2}{E}\frac{1}{\Delta t}$.

\begin{figure}[h]
\centering
\includegraphics[width=10cm]{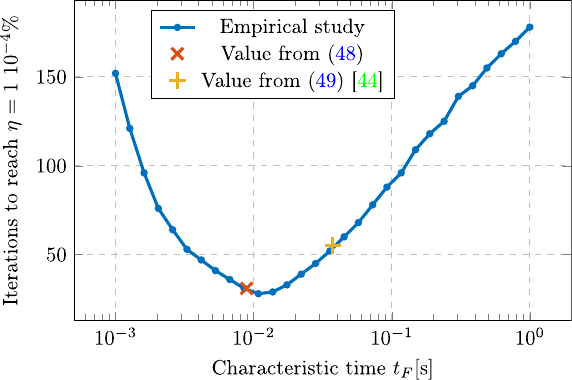}
\caption{Number of iterations depending on the value taken by $\tf$}
\label{fig:cv_study_1D_poroelasticity}
\end{figure}

The relevance of the analytical expression given in \eqref{eq:1D_poroelasticity_tf} is now compared to results provided by an empirical study. The study considers 30 different values of $\tf$, ranging from $\qty{1e-3}{\second}$ to $\qty{1e0}{\second}$. For each of these values, we have performed the LATIN algorithm (without PGD) and assessed the number of iterations to reach an error of $\qty{1e-4}{}\%$. We also ran the algorithm with the analytical values given in \cite{neron_2008} and \eqref{eq:1D_poroelasticity_tf} and assessed the number of iterations required to reach convergence for both values. The results are presented in \autoref{fig:cv_study_1D_poroelasticity}. For this study,  we use 60 bar elements for the spatial discretization and 100 times steps for the time discretization. Thus, the size of the time steps is $\Delta t = \qty{1.75}{\milli\second}$. All other parameters can be found in \cite{neron_2008}.

First, one can notice that the value of $\tf$ greatly influences the algorithm convergence rate. Moreover, we found that the analytical value \eqref{eq:1D_poroelasticity_tf} obtained by a dimensional analysis is almost equal to the optimal one. Thus, we validate the reasoning on a simple monodimensional example. We will now tackle the case of thermo-poroelasticity problems.

\subsubsection{Expression of the characteristic times for strongly nonlinear thermo-poroelasticity problems}\label{sec362}

We consider monodimensional examples and neglect the nonlinear terms to simplify the analysis. The reasoning remains the same as in poroelasticity. Thus, we start with the following system of equations:
\begin{equation}\label{eq:1D_thermoporoelasticity_continuous}
    \left\{
    \begin{aligned}
    &E\varepsilon - \couplP \p - \couplT\T = F(t)/S \\
    &\couplP \Dot{\varepsilon} + \invbiotmod \pdot -\couplTP \Tdot - \permeability \derivPart{^2 \p}{x^2} = 0 \\
    &\couplT \To \Dot{\varepsilon} - \couplTP \To \pdot + \genheatcapa \Tdot - \thermconductivity\derivPart{^2 \T}{x^2} = 0 
    \end{aligned}
    \right.
\end{equation}
Les champs solution $\p$ et $\T$ sont cherchés sous a forme d'un produit de fonctions spatiale et temporelle : 
We use the same notations as in \autoref{sec361} and introduce the notation $\ttheta$ for the characteristic time relative to the thermal part. As explained previously, one can write the pore pressure $\p$ and the temperature $\T$ with a product of an oscillating space function and an exponential time function:

\begin{equation}\label{eq:space-time_decomp_pressure_temperature}
    \left\{
    \begin{aligned}
    & \p(x,t) = f_p\left(\frac{x}{L_c}\right)g_p\left(\frac{t}{\tf}\right)\\
    & \T(x,t) = f_\theta\left(\frac{x}{L_c}\right)g_\theta\left(\frac{t}{\ttheta}\right)
    \end{aligned}
    \right.
\end{equation}
We express the strain either as $\Dot{\varepsilon} = \frac{\couplP}{E}\frac{\p}{\Delta t}$ or as $\Dot{\varepsilon} = \frac{\couplT}{E}\frac{\T}{\Delta t}$. Moreover, the forms of $\p$ and $\T$ given in \eqref{eq:space-time_decomp_pressure_temperature} are injected in the second and third equations of \eqref{eq:1D_thermoporoelasticity_continuous}:
\begin{equation}\label{eq:1D_thermoporoelasticity_approx}
    \left\{
    \begin{aligned}
    &\couplP \Dot{\varepsilon} - \invbiotmod \invtf \p  +\couplTP \invttheta \T - \permeability \left(-\frac{1}{L_c^2}\right)\p = 0 \\
    &\couplT \To \Dot{\varepsilon} + \couplTP \To \invtf\p - \genheatcapa \invttheta \T - \thermconductivity\left(-\frac{1}{L_c^2}\right)\T = 0 
    \end{aligned}
    \right.
\end{equation}
The two last equations of \eqref{eq:1D_thermoporoelasticity_approx} involve $\p$ and $\T$ that are replaceable thanks to the relation $\couplP \p = \couplT\T$ given by the first equation of \eqref{eq:1D_thermoporoelasticity_continuous}:
\begin{equation}\label{eq:1D_thermoporoelasticity_approx2}
    \left\{
    \begin{aligned}
    &\frac{\couplP^2}{E} \frac{\p}{\Delta t} - \invbiotmod \invtf \p  +\couplTP \frac{\couplP}{\couplT}\invttheta\p + \frac{\permeability}{L_c^2}\p = 0 \\
    &\To \frac{\couplT^2}{E}\frac{\T}{\Delta t}+ \couplTP \To \frac{\couplT}{\couplP}\invtf\T - \genheatcapa \invttheta \T + \frac{\thermconductivity}{L_c^2}\T = 0 
    \end{aligned}
    \right.
\end{equation}

Finally, we obtain the analytical expressions of the characteristic times $\tf$ and $\ttheta$ by solving the following linear system:
\begin{equation}\label{eq:1D_thermoporoelasticity_tf_ttheta}
\left\{
    \begin{aligned}
    & \invbiotmod \invtf - \couplTP \frac{\couplP}{\couplT}\invttheta = \frac{\permeability}{L_c^2} + \frac{\couplP ^2}{E}\frac{1}{\Delta t}\\
    & - \couplTP \To \frac{\couplT}{\couplP}\invtf + \genheatcapa \invttheta = \frac{\thermconductivity}{L_c^2} + \To\frac{\couplT ^2}{E}\frac{1}{\Delta t}
    \end{aligned}
\right.
\end{equation}

The relevance of the aforegiven expressions is assessed in \autoref{sec422}.

\subsection{Parametrized studies using the LATIN-PGD solver}\label{sec37}

Coming back to the purpose of this work, we now explain how to use the LATIN-PGD solver for parametrized studies. Here, we consider the case of variability in material parameters. \autoref{sec35} deals with using Proper Generalized Decomposition within the LATIN framework and shows how to get a reduced basis at the end of a computation. The strategy used to tackle parametrized studies in the frame of multiphysics problems relies on two major concepts formulated in \cite{wurtzer_2024}.

The first key point of the strategy consists of reusing the basis built along a computation and progressively enriching it along the computations. We sensibly suppose that a solution does not change much with a slight modification of the material parameters, and we reduce the number of modes generated if a basis already exists at the beginning of the algorithm. The second key point of the method lies in the algorithm's initialization. Instead of starting with an elastic initialization, starting the algorithm with a computed solution is possible. However, choosing a relevant computed solution for the initialization is necessary. It has been shown in \cite{boucard_2003} that in the frame of material variability with quasi-static loadings, a small distance in the parametric space leads to a small distance between the corresponding responses. Therefore, we initialize a computation with the closest computed solution in the parametric space.

\begin{figure}[h]
    \centering
    \includegraphics[width = 12cm]{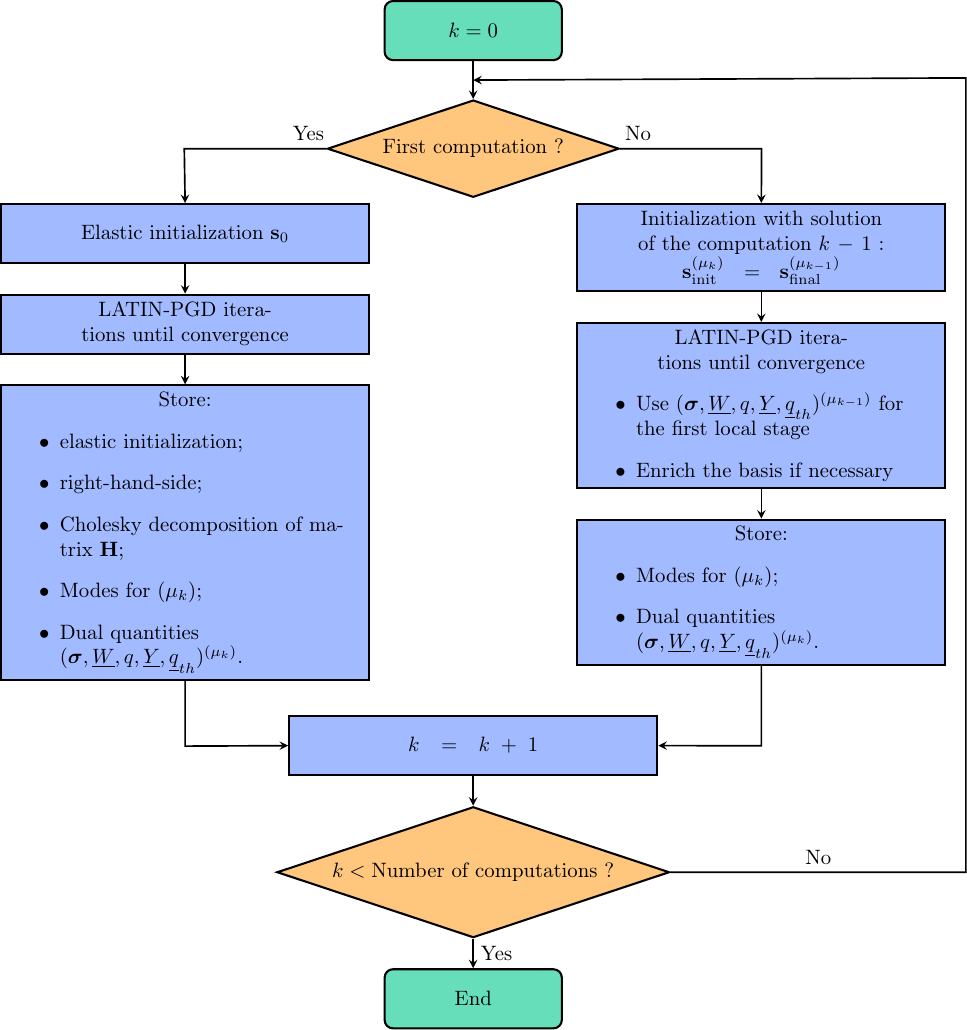}
    \caption{Flowchart for a parametric study in thermo-poroelasticity using the LATIN-PGD solver}\label{fig:flowchart_parametrized}
\end{figure}

Considering the abovementioned comments, we cross the parametric space step-by-step and use the last computed solution for the initialization. We summarize the strategy with the flowchart in \autoref{fig:flowchart_parametrized}.

\section{Results}\label{sec4}

This section focuses on the results provided by the monolithic and LATIN-PGD solver for two nonlinear thermo-poroelasticity problems. First, both solvers are validated on an academic example for which reference results are provided in the literature. Then, we apply the method to an industrial test case modeling a ceramic membrane used for filtering fluids. For both examples, we undertake parametrized studies on different material properties.

\subsection{An academic test case: Aboustit problem}\label{sec41}

We first focus on a common benchmark used in thermo-poroelasticity. The Aboustit problem was first introduced in \cite{aboustit_1985} and consists of a one-dimensional thermoelastic consolidation problem. We study a column of soil of length $L=\qty{7}{\meter}$ and cross section $S =\qty{2}{\meter}\times\qty{2}{\meter}$ under a constant mechanical loading $\sigma_z$ and a constant thermal loading $\Td$ on its top surface. \autoref{fig:scheme_aboustit} displays the boundary conditions. The displacements are constrained along $\vect{z}$ for the bottom surface and along $\vect{x}$ (respectively $\vect{y}$) for the lateral surfaces with normal $\pm \vect{x}$ (respectively $\pm \vect{y}$). For fluid physics, the pore pressure equals zero on the top surface, and the fluid flux is null on the lateral and bottom surfaces. The thermal flux is null on the lateral surfaces and the bottom. The numerical values of the boundary conditions and the initial conditions are given in \autoref{tab:bc_ic_aboustit}.

\begin{minipage}[t]{0.4\textwidth}
    \vspace{0pt}
    \centering
    \includegraphics[width=\textwidth]{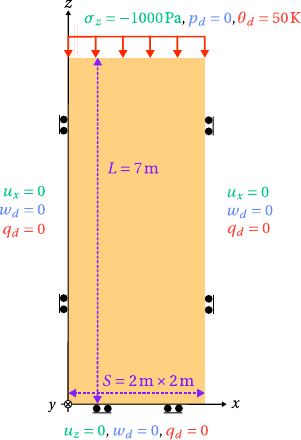}
    \captionof{figure}{Representation of Aboustit problem}
    \label{fig:scheme_aboustit}
\end{minipage}
\hfill
\begin{minipage}[t]{0.55\textwidth}
    \vspace{40pt}
    \centering
    \captionof{table}{Boundary and initial conditions for Aboustit problem}
    \label{tab:bc_ic_aboustit}
    \begin{tabular}{c c}
    \toprule
    Parameters & Value (SI units)\\
    \hline
        Traction force $\sigma_z$ & $\qty{1000}{\pascal}$ \\
        Temperature & $\qty{50}{\kelvin}$  \\
        increment $\Td$ &    \\
        Initial pore & $\qty{1000}{\pascal}$  \\
        pressure $\po$ & \\
        Initial temperature & $\qty{0}{\kelvin}$ \\
        increment $\theta_0$ &  \\
        Initial & $\qty{273}{\kelvin}$ \\
        temperature $\To$ &  \\
    \bottomrule
    \end{tabular}
\end{minipage}
The material properties of the problem are recapitulated in \autoref{tab:para_mat_aboustit}. To avoid conditioning issues, we introduce a new unit system such that:
\begin{equation*}
    M = \qty{1e6}{\kilogram}, \: S = \qty{1}{\second}, \: L = \qty{1}{\meter}, \: T = \qty{1}{\kelvin}
\end{equation*}

\begin{table}[h]
    \centering
    \caption{Material parameters for Aboustit problem}
    \label{tab:para_mat_aboustit}
    \begin{tabular}{c c c}
    \toprule
    Parameters & Value (SI units) & Value (new units) \\
    \hline
        Young modulus $E$ & $\qty{6}{\mega\pascal}$ & $\qty{6}{M \: L^{-1} \:S^{-2}}$  \\
        Poisson coefficient $\nu$ & $0.4$ & $0.4$ \\
        Porosity $\porosity$ & $0.2$ & $0.2$ \\
        Fluid dynamic viscosity $\fluidvisco$ & $\qty{1e-3}{\pascal\s}$ &$\qty{1e-9}{M \: L^{-1} \:S^{-1}}$\\
        Heat generalized capacity $\genheatcapa$ & $\qty{167}{\kilo\joule\per\cubic\meter\per\kelvin}$ & $\qty{167e-3}{M\: L^{-1}\: S^{-2}\: T^{-1}}$\\
        Thermal conductivity $\thermconductivity$ & $\qty{837}{\watt\per\meter\per\kelvin}$ & $\qty{837e-6}{M\: L\: S^{-3}\: T^{-1}}$\\
        Skeleton expansion coefficient $\expcoefsol$ & $\qty{3e-7}{\per\kelvin}$ & $\qty{3e-7}{T^{-1}}$\\
        Porous medium expansion coefficient $\expcoef$ & $\qty{1.2e-5}{\per\kelvin}$ & $\qty{1.2e-5}{T^{-1}}$\\
    \bottomrule
    \end{tabular}
\end{table}

Some material parameters have non-physical values, such as the Young modulus. Indeed, the Aboustit problem is a benchmark used to validate a thermo-poroelasticity problem and does not aim to model a physical consolidation problem. Moreover, as noticed in \cite{pogacnik_2011}, some material parameters are lacking, or there is an uncertainty on their value. For instance, the hydraulic conductivity is given by $\hydconductivity = \qty{4e-6}{\meter\per\hour} = \qty{1.1e-9}{\meter\per\second}$ in \cite{lewis_schrefler_1999} while $\hydconductivity = \qty{3.92e-5}{\meter\per\second}$ in \cite{cui_2018} and $\hydconductivity = \qty{4.6e-17}{\meter\per\second}$ in \cite{sanavia_2008}. We undertake an empirical study to find the material parameters fitting with the results from the literature. With such a study, we found that a Biot modulus of $\biotmod = \qty{4.6e-2}{\mega\pascal}$ and an intrinsic permeability $\intpermeability = \qty{1.29e-3}{\square\meter}$ enables to recover the results given in \cite{lewis_schrefler_1999}. Thus, the material parameters given in \autoref{tab:para_mat_aboustit} can be completed with the values given in \autoref{tab:add_para_mat_aboustit}.
\begin{table}[h]
    \centering
    \caption{Additional parameters for Aboustit problem found with an empirical study and relations given in \autoref{sec21}}
    \label{tab:add_para_mat_aboustit}
    \begin{tabular}{c c c}
    \toprule
    Parameters & Value (SI units) & Value (new units) \\
    \hline
        Biot modulus $\biotmod$ & $\qty{46}{\kilo\pascal}$ & $\qty{4.6e-2}{M \: L^{-1} \:S^{-2}}$ \\
        Biot coefficient $\couplP$ & $1$ & $1$\\
        Intrinsic permeability $\intpermeability$ & $\qty{1.29e-3}{\square\meter}$ & $\qty{1.29e3}{L^2}$\\
        Permeability $\permeability$ & $\qty{1.29e-6}{\cubic\meter\second\per\kilogram}$ & $\qty{1.29}{L^{3}\: S\: M^{-1}}$\\
    \bottomrule
    \end{tabular}
\end{table}

For the Aboustit problem, we consider a simplified equation for the generalized heat source \cite{lewis_schrefler_1999}:
\begin{equation}\label{eq:aboustit_heat}
    R = \genheatcapa \Tdot - \densityflu \heatflu \W \cdot \X
\end{equation}
The simplification of \eqref{eq:generalized_heat_law} leads to canceling $\Mtu{C}$ and $\Mtp{C}$ in the monolithic system \eqref{eq:matrix_system_mono} and all related coupling terms in the last equation \eqref{eq:coupled5} of the differential equations system in the LATIN solver.

\subsubsection{Validation of the LATIN-PGD solver}\label{sec411}

We first compare the solution given by the linear monolithic solver to the results from the literature. For the spatial discretization, we use composite elements whose order is one for the fluid pressure and the temperature and two for the displacement \cite{sandhu_1977,aboustit_1982}. \autoref{tab:time_discretization_Aboustit} provides the time discretization, inspired by the one given in \cite{lewis_schrefler_1999}.
\begin{table}[h]
    \centering
    \caption{Time discretization used for Aboustit problem}
    \begin{tabular}{c c}
    \toprule
        Size of time steps & Number of time steps \\
        \hline 
        $\qty{5}{\milli\second}$ & 20\\
        $\qty{50}{\milli\second}$ & 20\\
        $\qty{5}{\second}$ & 20\\
        $\qty{50}{\second}$ & 20\\
        $\qty{500}{\second}$ & 40\\
    \bottomrule
    \end{tabular}    
    \label{tab:time_discretization_Aboustit}
\end{table}

\begin{figure}[h]
    \centering
    \begin{subfigure}{0.32\textwidth}
        \includegraphics[width=4.25cm]{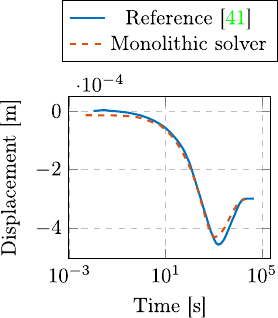}
        \caption{Displacement at $z=\qty{7}{\meter}$}
    \end{subfigure}
    \hfill
    \begin{subfigure}{0.32\textwidth}
        \includegraphics[width=4.25cm]{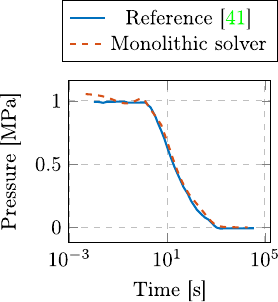}
        \caption{Pore pressure at $z=\qty{6}{\meter}$}
    \end{subfigure}
    \hfill
    \begin{subfigure}{0.32\textwidth}
        \includegraphics[width=4.25cm]{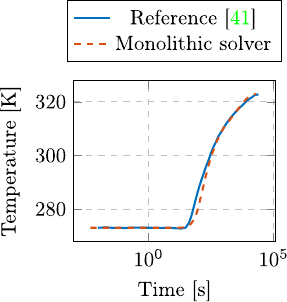}
        \caption{Temperature at $z=\qty{6}{\meter}$}
    \end{subfigure}
    \caption{Comparison between solutions given by the literature \cite{lewis_schrefler_1999} and the monolithic solver}
    \label{fig:comp_litt}
\end{figure}

\autoref{fig:comp_litt} compares the results given by the monolithic solver with the plots of each primal field given in \cite{lewis_schrefler_1999}. First, we notice that we recover the same trends for all physics. We observe a slight difference between the reference and the solution given by the monolithic solver for the displacement field and the fluid pressure one. Such observation is explainable by the uncertainties of the material parameters. It could also be due to the nonlinear effects neglected within the monolithic solver. The plots for the temperature field are almost coincident, as all material parameters were provided in the literature. Finally, one can observe oscillations in the pressure field at first time steps. As explained in \cite{cui_2018}, those oscillations are due to the time discretization chosen.

\begin{figure}[h]
    \centering
    \begin{subfigure}{0.49\textwidth}
        \includegraphics[width=7.5cm]{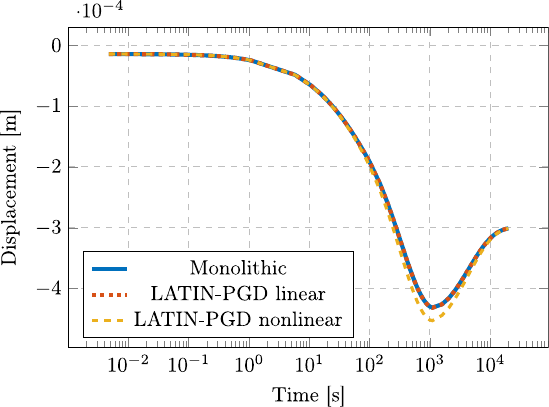}
        \caption{Displacement at $z=\qty{7}{\meter}$}
    \end{subfigure}
    \hfill
    \begin{subfigure}{0.49\textwidth}
        \includegraphics[width=7.5cm]{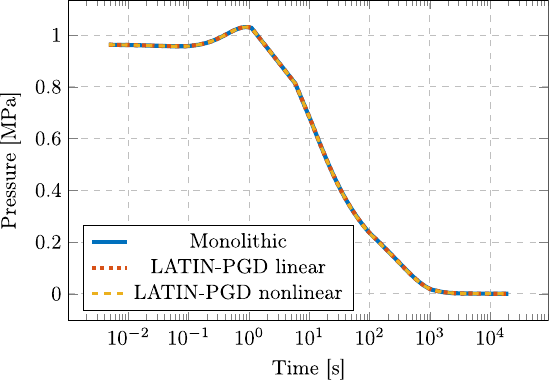}
        \caption{Pore pressure at $z=\qty{6}{\meter}$}
    \end{subfigure}
    \hfill
    \begin{subfigure}{0.49\textwidth}
        \includegraphics[width=7.5cm]{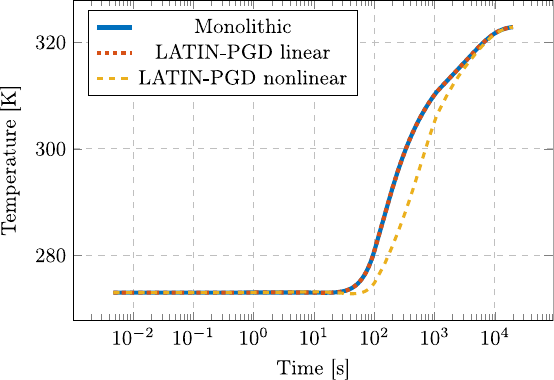}
        \caption{Temperature at $z=\qty{6}{\meter}$}
    \end{subfigure}
    \caption{Comparison along time between solutions given by the different solvers}
    \label{fig:comp_all}
\end{figure}

Now that the monolithic solver has been validated based on the results provided in the literature, we can compare the results from the monolithic solver and the LATIN-PGD one. \autoref{fig:comp_all} displays the results given by the monolithic solver and the LATIN-PGD solver with a convergence threshold $\eta_c = \qty{1e-3}{}$. It is visible that the results provided by the linear LATIN-PGD solver are consistent with those of the reference monolithic solution. Moreover, we observe minor differences due to the presence of nonlinearities.

\subsubsection{Building independent bases with PGD}\label{sec412}

\begin{figure}[h]
    \centering
    \begin{subfigure}{0.49\textwidth}
      \includegraphics[width=7cm]{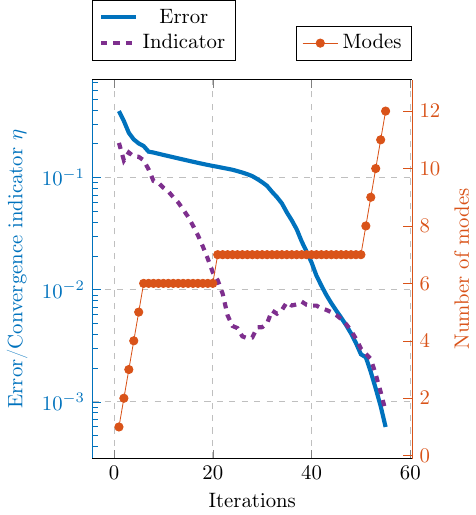}
      \caption{Solid}
    \end{subfigure}
    \hfill
    \begin{subfigure}{0.49\textwidth}
      \includegraphics[width=7cm]{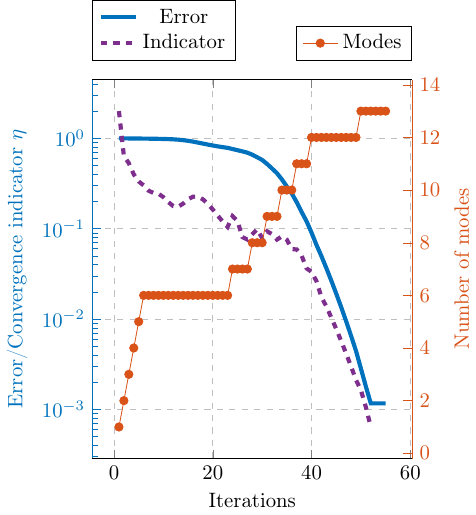}
      \caption{Fluid}
    \end{subfigure}
    \hfill
    \begin{subfigure}{0.49\textwidth}
      \includegraphics[width=7cm]{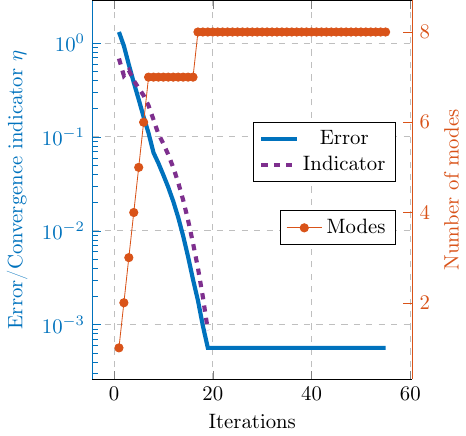}
      \caption{Thermics}
    \end{subfigure}
    \caption{Error with respect to monolithic reference and convergence indicator along iterations in opposition with the number of modes for the Aboustit problem}
    \label{fig:err_modes_iter_aboustit}
\end{figure}

\autoref{fig:err_modes_iter_aboustit} illustrates the evolution of the error and the convergence indicator along iterations in the linear frame. The number of modes added along iterations is also plotted on the right axis. First, we notice that the error and the indicator follow the same trends, which is reassuring for the case where one does not own any reference solution. Secondly, one observes that the basis size differs for all physics, which is one advantage of the PGD. We also remark that the number of modes is affordable despite the relatively complex time discretization involving significant differences in the time step sizes.

\autoref{tab:modes_lin_nonlin_aboustit} shows the number of modes to decompose each primal field in the linear and nonlinear cases with a convergence threshold $\eta_c = \qty{1e-3}{}$. We observe that the performances of the LATIN-PGD solver do not deteriorate although the problem introduces nonlinearities. The number of modes is higher to accurately decompose the nonlinear temperature field but remains stable for solid and fluid physics. This point is consistent with the fact that the nonlinearities are located in the generalized heat source expression and thus influence mostly the thermal part of the problem. Moreover, the linear computation lasts $\qty{19.53}{\second}$ while the nonlinear one lasts $\qty{19.58}{\second}$. The ratio of both computation times is only $1.003$, proving the efficiency of the LATIN-PGD solver in dealing with nonlinearities in coupled thermo-poroelasticity problems.
\begin{table}[h]
    \caption{Number of modes at the end of linear and nonlinear computations}
    \label{tab:modes_lin_nonlin_aboustit}
    \centering
    \begin{tabular}{c c c}
    \toprule
     & Linear & Nonlinear \\
     \hline
    Displacement & 12  & 12 \\
    Fluid pressure & 13 & 12\\
    Temperature & 8 & 15 \\
    \bottomrule
    \end{tabular}
\end{table}

\subsubsection{Variation of the thermal expansion coefficient of the solid phase}\label{sec413}

\begin{figure}[h]
    \centering
    \includegraphics[width=12cm]{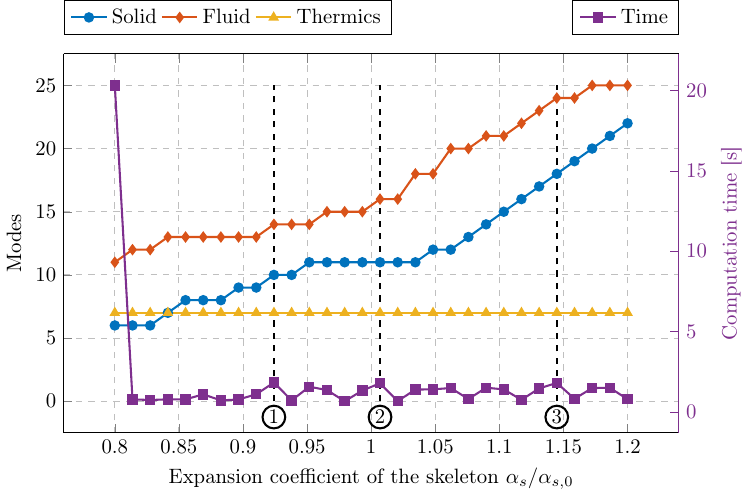}
    \caption{Results of parametrized study on $\expcoefsol$ with the Aboustit problem}
    \label{fig:aboustit_parametrized}
\end{figure}

One can now perform several computations and take advantage of the LATIN-PGD method. In this example, we consider a variability in the thermal expansion coefficient of the solid phase $\expcoefsol$. As explained in \autoref{sec2}, the thermal expansion coefficient of the porous media $\expcoef$ and the coupling coefficient between solid and thermal parts $\couplT$ are directly linked to $\expcoefsol$. Thus, a change in $\expcoefsol$ should involve changes in the behavior of all physics. In the case of the Aboustit problem, the solid and fluid physics does not influence the thermal part, as shown in \eqref{eq:aboustit_heat}. Thus, one should observe changes only for the solid and fluid parts. We choose to perform 30 computations, for which $\expcoefsol$ is varying in the uniformly divided parametric space $\mathcal{D}_{\expcoefsol} = [0.8, 1.2]\cdot\expcoefsolinit$ where $\expcoefsolinit$ designates the thermal expansion coefficient of the solid phase given in \autoref{tab:para_mat_aboustit}. The parametric space is crossed for growing $\expcoefsol$. \autoref{fig:aboustit_parametrized} shows the results of the parametrized study concerning the computation times and the number of added modes.

\autoref{fig:aboustit_parametrized} depicts the number of modes obtained at the end of each computation for the three physics. First, we recover that no new mode is added in thermics, which is consistent with the equations of the Aboustit problem.  We also notice that for several computations, no new mode is added for the solid and fluid physics, and the solution corresponding to the new value of $\expcoefsol$ only requires updates of the time functions to be relevant. Considering that the updates of time functions are cheap compared to generating a new mode, the PGD algorithm is well-adapted to parametric studies. On the right axis of \autoref{fig:aboustit_parametrized}, one can read the computation time of each new simulation. The parametric space is crossed with growing $\expcoefsol$, which means that $\expcoefsol = 0.8\expcoefsolinit$ corresponds to the first simulation initialized without any basis and with a decoupled elastic solution. The first computation is indeed obviously slower than the others. The first computation lasts $\qty{20.31}{\second}$ while all others last in average $\qty{1.157}{\second}$. However, one notes that some of the other computations are longer than others. For instance, the computations at points $\circled{1}$, $\circled{2}$ and $\circled{3}$ respectively last $\qty{1.837}{\second}$, $\qty{1.789}{\second}$ and $\qty{1.812}{\second}$. \autoref{fig:aboustit_parametrized} shows these points correspond to computations during which several modes have been generated, consistent with the more expensive cost of the modes generation stage. The simulation of 30 problems lasts $\qty{53.87}{\second}$. A single computation for $\expcoefsol = 0.8\expcoefsolinit$ lasts $\qty{20.31}{\second}$ and $\qty{20.79}{\second}$ for $\expcoefsol = 1.2\expcoefsolinit$. Therefore, we consider that a single computation for $\expcoefsol \in \mathcal{D}_{\expcoefsol}$ last in average $\tau_{single} = \qty{20.55}{\second}$. We define the gain $G$ in computation time as the ratio between the computation time for the naive approach denoted $\tau_{naive}$ and the computation time with the LATIN-PGD approach denoted $\tau_{LATIN}$. The gain of the method in the studied case is then given by:
\begin{equation}\label{eq:gain_aboustit}
    G = \frac{\tau_{naive}}{\tau_{LATIN}} \simeq \frac{30\tau_{single}}{\tau_{LATIN}}=11.4
\end{equation}
To conclude, the strategy employed here is much more adapted than a naive approach, consisting of performing all simulations sequentially without accounting for the similarities between computations.

\subsection{Application to a ceramic membrane}\label{sec42}

\begin{figure}[h]
    \centering
    \begin{subfigure}{0.49\textwidth}
      \centering
      \includegraphics[width=5cm]{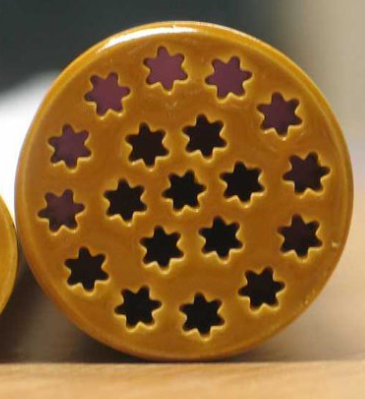}
      \caption{Picture taken from the \textit{Star-Sep}\texttrademark\ datasheet}
      \label{fig:datasheet_star_sep}
    \end{subfigure}
    \hfill
    \begin{subfigure}{0.49\textwidth}
      \centering
      \includegraphics[width=7.5cm]{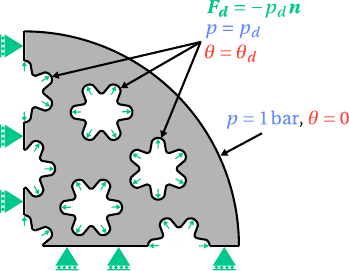}
      \caption{Representation of the geometry used for the numerical study with boundary conditions}
      \label{fig:scheme_star_sep}
    \end{subfigure}
    \caption{The \textit{Star-Sep}\texttrademark\ technology developed by \textit{Mantec} for ceramic membranes}
    \label{fig:star_sep_mantec}
\end{figure}

We now consider the more complex case of ceramic membranes used to filter fluids. Those membranes are made of ceramics, which are porous materials. With the pressure, the fluid crosses the ceramics, and bacteria are retained in the pores. Recent developments have led to various designs optimizing the filtration area. Among several types of ceramic membranes, we choose to study the \textit{Star-Sep}\texttrademark\ technology offered by the company \textit{Mantec}. The specific shape of the channels enables the increase of the filtration area, the reduction of the volume of the material, and the reduction of the pumping energy. We focus here on a ceramic membrane composed of 19 channels shown in \autoref{fig:datasheet_star_sep}. For the numerical study, we restrain the study to a quarter of the membrane as presented in \autoref{fig:scheme_star_sep}. The geometric parameters originate from the \textit{Mantec} datasheet and are summed up in \autoref{tab:para_geo_star_sep}. The stars are uniformly shared around the membrane center.

\begin{table}[h]
    \caption{Geometric parameters of the \textit{Star-Sep}\texttrademark\ membrane with 19 channels}
    \label{tab:para_geo_star_sep}
    \centering
    \begin{tabular}{c c}
    \toprule
    Parameters & Numerical value \\
    \hline
    Membrane diameter & $\qty{32}{\milli\meter}$ \\
    Radial distance between two stars & $\qty{5.76}{\milli\meter}$ \\
    Outer star diameter & $\qty{4.6}{\milli\meter}$ \\
    Inner star diameter & $\qty{2.8}{\milli\meter}$ \\
    Radius of curvature of the star internal fillet & $\qty{0.3}{\milli\meter}$\\
    Radius of curvature of the star external fillet & $\qty{0.54}{\milli\meter}$\\
    \bottomrule
    \end{tabular}
\end{table}

The ceramic membrane normal displacements are constrained on the symmetric planes, as depicted in \autoref{fig:scheme_star_sep}. The fluid crossing the membrane imposes a pore pressure in the channels, generating a mechanical loading on the inner surface of the channels. The pore pressure is uniformly increasing from $\qty{0}{bar}$ to $\pd ~= ~ \qty{8}{bar} ~=~ \qty{1.14e-3}{M \: L^{-1} \:S^{-2}}$ during the first $\qty{12.5}{\milli\second}$ and is then constant and equal to $\pd$. A thermal loading is also present in the channels and given by a ramp increasing from $\qty{0}{\kelvin}$ to $\Td  ~=~ \qty{80}{\kelvin} = \qty{2.73e-1}{T}$. The outer boundary of the membrane is at the atmospheric pressure and an ambient temperature $\To = \qty{293}{\kelvin} = \qty{1}{T}$.

The material parameters refer to a porous ceramic material and are all given in \autoref{tab:para_mat_star_sep}. One can notice that the different parameters differ by several orders of magnitude in the international system of units. Thus, we define a new unit system given in \cite{neron_2007}:
\begin{equation*}
    M = \qty{7e8}{\kilogram}, \: S = \qty{1}{\second}, \: L = \qty{1}{\meter}, \: T = \qty{293}{\kelvin}
\end{equation*}

The time domain is discretized uniformly in 120 time steps, and the final time step is equal to $T_f = \qty{25}{\milli\second}$. The spatial domain is discretized in 52,630 elements, resulting in 65,442 nodes for the solid part (Tet10 elements) and 10,019 nodes for the fluid and thermal parts (Tet4 elements). We consider here a strong coupling between all physics. Therefore, unlike the Aboustit problem, we consider the equations given in \eqref{eq:strong_formulation_TPM} with all coupling terms.

\subsubsection{Solutions obtained for the primal fields}\label{sec422}

The choice of the search directions has been developed in \autoref{sec362}. Using numerical values given in \autoref{tab:para_mat_star_sep}, the search directions corresponding to the present problem are given by:
\begin{equation}\label{eq:first_values_tf_ttheta}
\left\{
    \begin{aligned}
    & \invtf = \qty{6.31e3}{\per\second}\\
    & \invttheta = \qty{1.68e2}{\per\second}
    \end{aligned}
\right.
\end{equation}
When performing an empirical study, one finds that the convergence speed is optimal with:
\begin{equation}\label{eq:optim_values_tf_ttheta}
\left\{
    \begin{aligned}
    & \invtf = \qty{4.55e2}{\per\second}\\
    & \invttheta = \qty{1.70e2}{\per\second}
    \end{aligned}
\right.
\end{equation}
While the value of $\ttheta$ is well predicted, the values of $\tf$ obtained analytically and empirically differ by a factor of 14. This difference may be explained by assumptions that are not all respected. The nonlinear terms have been neglected, and the approximation of the space derivative might be inappropriate considering the complexity of the geometry. Although the reasoning still requires improvements, it provides a preliminary indication of the search directions.

\begin{figure}[h]
    \centering
    \includegraphics[width=15cm]{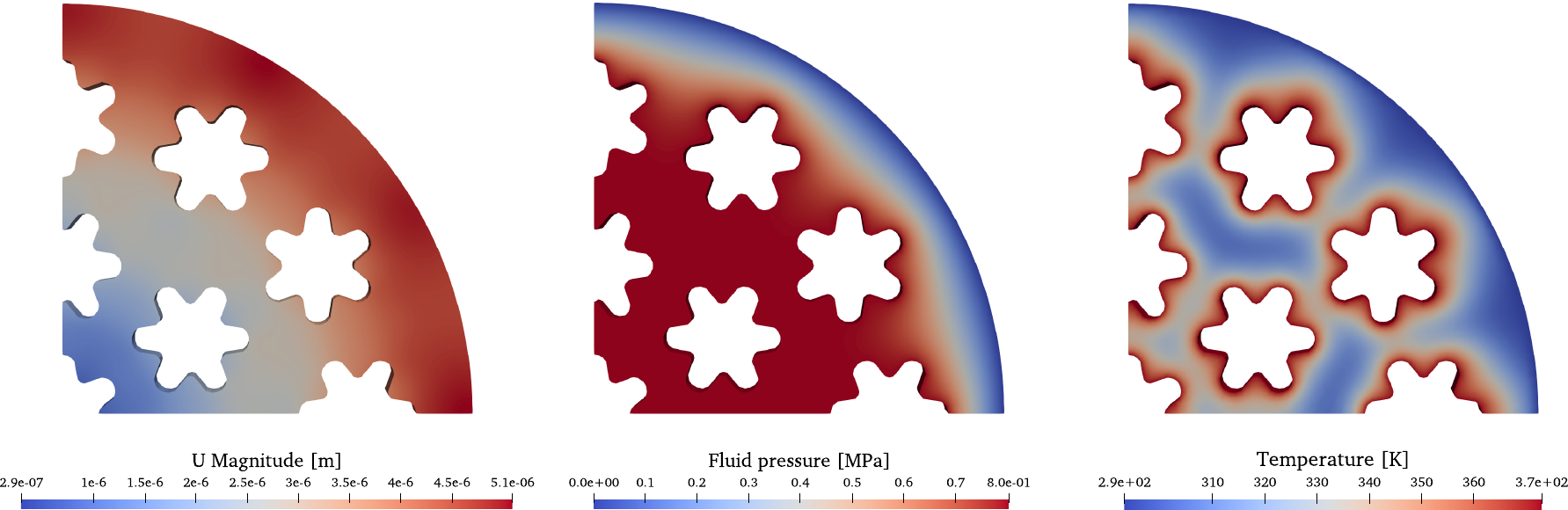}
    \caption{Primal fields of the solid, fluid and thermal physics at $t = \qty{25}{\milli\second}$}
    \label{fig:primal_fields_star_sep_NL}
\end{figure}

The results obtained with the LATIN-PGD solver ($\eta_c = \qty{1e-3}{}$) for the three primal fields are depicted in \autoref{fig:primal_fields_star_sep_NL}. The primal fields are decomposed in a space-time form with:
\begin{itemize}
    \item 15 modes for the displacement;
    \item 6 modes for the pore pressure;
    \item 18 modes for the temperature.
\end{itemize}

There are variations in temperature distribution from one channel to another. The area with maximum temperatures is wider at the periphery of the membrane, where the temperature gradient is the highest. As the nonlinear terms involve the temperature gradient, the appearance of the temperature field is consistent.

\begin{figure}[h]
    \centering
    \begin{subfigure}{0.49\textwidth}
        \centering
      \includegraphics[width=4cm]{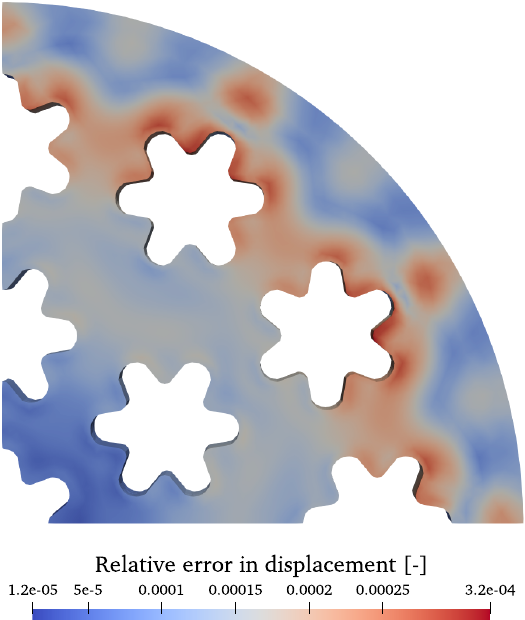}
      \caption{Displacement}
      \label{fig:starsep_error_disp}
    \end{subfigure}
    \hfill
    \begin{subfigure}{0.49\textwidth}
        \centering
      \includegraphics[width=4cm]{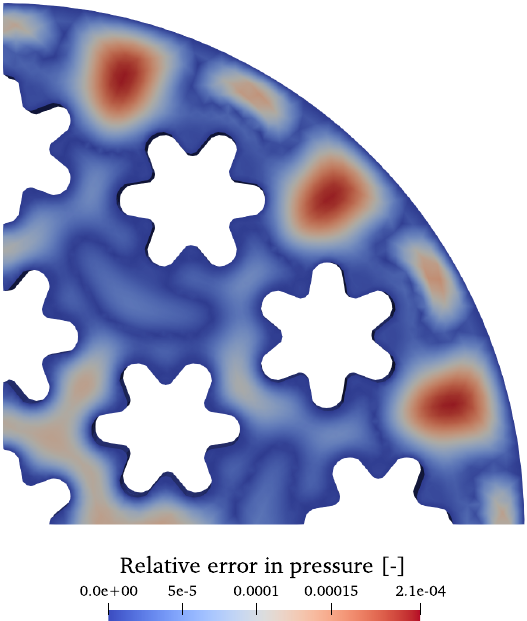}
      \caption{Pressure}
      \label{fig:starsep_error_press}
    \end{subfigure}
    \begin{subfigure}{0.49\textwidth}
        \centering
      \includegraphics[width=4cm]{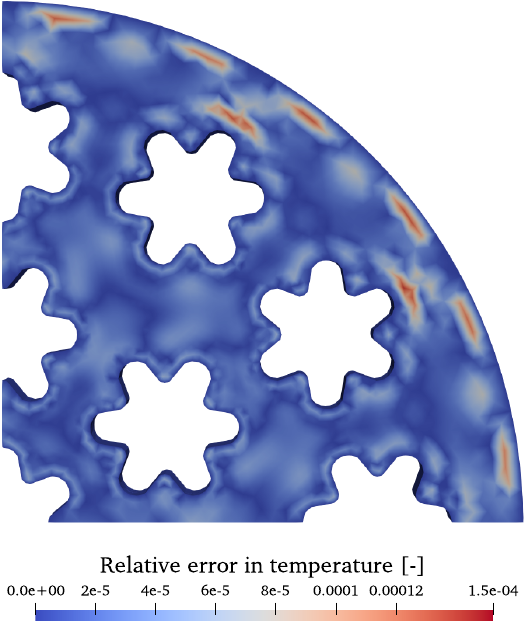}
      \caption{Temperature}
      \label{fig:starsep_error_temp}
    \end{subfigure}
    \hfill
    \begin{subfigure}{0.49\textwidth}
        \centering
      \includegraphics[width=7cm]{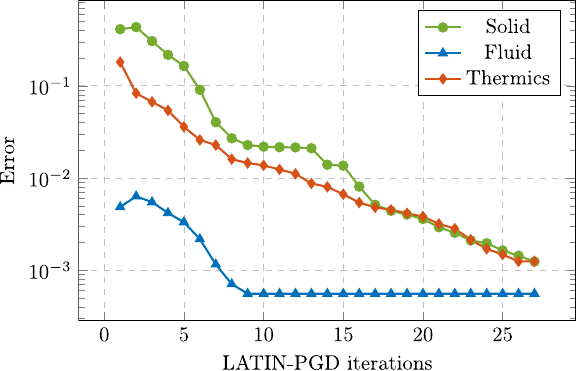}
      \caption{Error evolution along LATIN iterations}
      \label{fig:starsep_error_plot}
    \end{subfigure}
    \caption{Relative errors between the reduced-order and full-order solutions at $t = \qty{25}{\milli\second}$ and evolution of the error along LATIN-PGD iterations}
    \label{fig:starsep_error_NL_PGD}
\end{figure}

The reduced-order solution is compared to a reference computed with the full-order LATIN solver and with a low tolerance threshold for the convergence indicator. \autoref{fig:starsep_error_disp}, \autoref{fig:starsep_error_press}, and \autoref{fig:starsep_error_temp} respectively depict the error towards the reference solution fields for the displacement, the pore pressure and the temperature at $t = \qty{25}{\milli\second}$. The relative error between the two solutions is plotted on \autoref{fig:starsep_error_plot}. One notices that at the end of the simulation, the error reaches $\qty{1e-3}{}$ for each physics.

\begin{figure}[h]
    \centering
    \begin{subfigure}{0.49\textwidth}
        \centering
      \includegraphics[width=5cm]{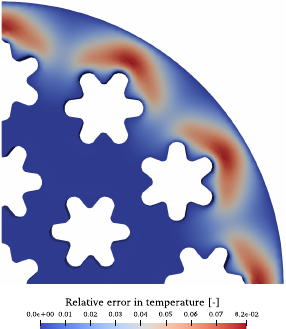}
      \caption{Temperature}
      \label{fig:starsep_rel_error_temp}
    \end{subfigure}
    \hfill
    \begin{subfigure}{0.49\textwidth}
        \centering
      \includegraphics[width=5cm]{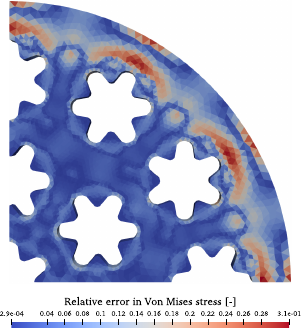}
      \caption{Von Mises stress}
      \label{fig:starsep_rel_error_vm}
    \end{subfigure}
    \caption{Relative error between the linear and the nonlinear solutions for two different fields at $t = \qty{25}{\milli\second}$}
    \label{fig:starsep_rel_error_lin_NL}
\end{figure}

On \autoref{fig:starsep_rel_error_lin_NL}, we plot the map of the relative error between the linear and nonlinear reduced-order solutions. One gets the relative error by dividing the difference of the two fields by the maximal value of the nonlinear field. \autoref{fig:starsep_rel_error_temp} shows the relative error on the temperature. We verify that the error is maximal on the periphery of the membrane. High temperatures may induce significant differences in the mechanical stresses. \autoref{fig:starsep_rel_error_vm} depicts the relative error on the Von Mises stress field. We notice a maximum difference of $30 \%$ between the linear and the nonlinear fields. Therefore, the nonlinear terms are not negligible and must be considered when choosing an appropriate material for the ceramic membrane. Finally, the complexity of the solution due to the nonlinearities induces an increase in the number of iterations before reaching convergence. With a tolerance threshold $\eta = \qty{1e-3}{}$, the reduced-order solution is obtained in 28 iterations in the nonlinear case and in 15 iterations in the linear one.

\subsubsection{Variation of three material parameters}\label{sec423}

\begin{figure}[h]
    \centering
    \includegraphics[width=12cm]{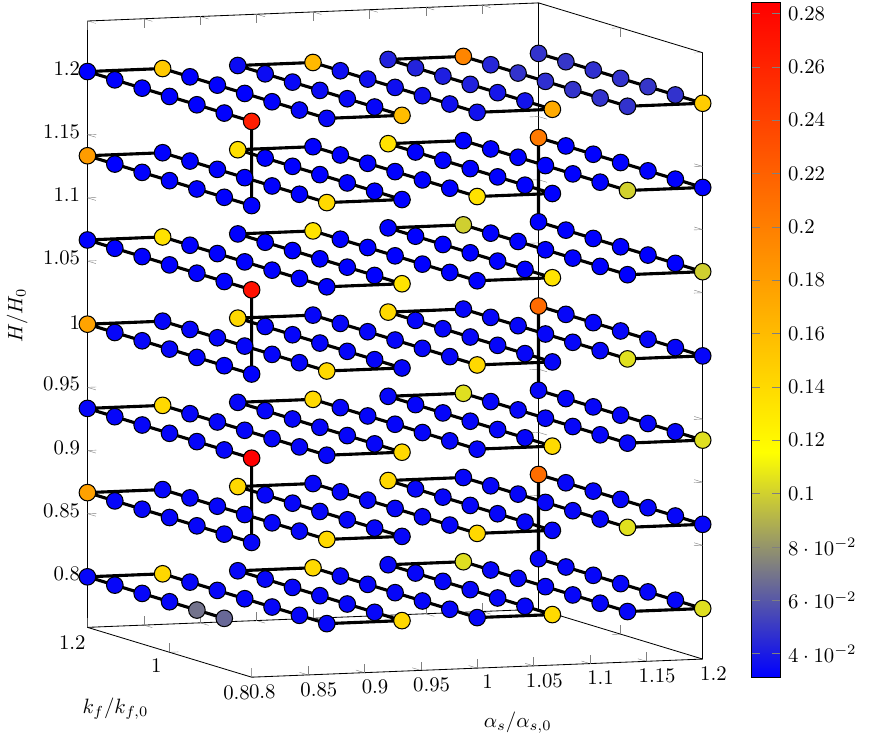}
    \caption{Normalized computation times of each computation represented in the parametric space}
    \label{fig:normalized_times_Star_Sep}
\end{figure}

A parametric study on several material parameters has been undertaken with variabilities on both the fluid and solid parts. The ceramic membrane is used to filter different fluids, which means that one should find a ceramic material suitable for different filtering configurations. We choose to perform a parametric study on the thermal conductivity of the fluid $\thermconductivityflu$, the thermal expansion coefficient of the solid $\expcoefsol$, and the permeability $\permeability$. The parametric space is given by $\mathcal{D} = \left\{[0.8, 1.2]\cdot\thermconductivityfluinit \cup [0.8, 1.2]\cdot\expcoefsolinit \cup [0.8, 1.2]\cdot\permeabilityinit\right\}$. For each parameter, 7 uniformly divided values are tested, resulting in $7^3 = 343$ sets of parameters. The parametric space is crossed step-by-step from the point $(0.8\expcoefsolinit, 0.8\thermconductivityfluinit, 0.8\permeabilityinit)$ to the point $(1.2\expcoefsolinit, 1.2\thermconductivityfluinit, 1.2\permeabilityinit)$. The path followed in the parametric space is depicted by the black line on \autoref{fig:normalized_times_Star_Sep} and enables the change of only one parameter from a computation to another. As shown in \cite{neron_2015}, choosing a random order of the parameters sets results in higher computation times. Other methods exist for the rational exploration of the parametric space. While \cite{heyberger_2013} presents a method inspired by the Reduced Basis method, an optimal path can also be defined using a particular proximity indicator \cite{daby_2025}.\\
We first focus on the computation times. \autoref{fig:normalized_times_Star_Sep} depicts the normalized times of each computation in the parametric space. The normalized time of a computation is the ratio between its simulation time and the first computation time. One notes that all computations are significantly faster than the first: they last between $4\%$ and $28\%$ of the first computation time. Furthermore, one observes that simulations are longer when a change in $\expcoefsol$ or $\permeability$ occurs, which means that the thermal conductivity of the fluid has a negligible influence on the solution. To compute the gain in time of the method, we measure the simulation times of the computation times at each extremum of the parametric space. The average simulation time of those 8 computations is equal to $\tau_{single} = \qty{1.11e3}{\second}$. The simulation of the 343 computations using the presented approach for parametrized problems is equal to $\tau_{LATIN} = \qty{1.86e4}{\second}$. Thus, the gain of the method in this example is given by:
\begin{equation*}
     G = \frac{\tau_{naive}}{\tau_{LATIN}} \simeq \frac{343\tau_{single}}{\tau_{LATIN}}=20.35
\end{equation*}

\begin{figure}[h]
    \centering
    \includegraphics[width=12cm]{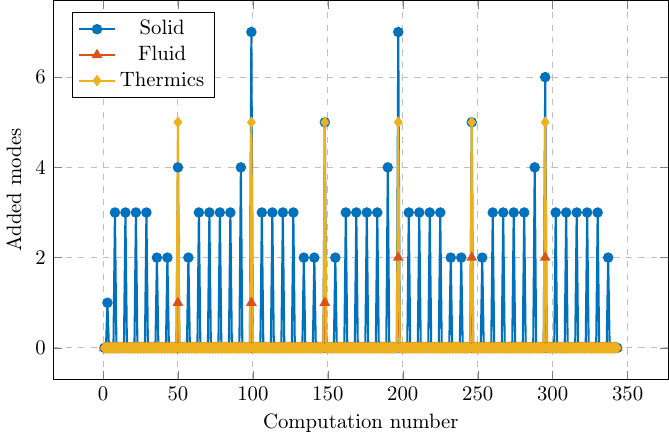}
    \caption{Number of added modes per physics along computation}
    \label{fig:modes_multiparametric_Star_Sep}
\end{figure}

We now look at the evolution of the number of modes along computations. The first computation leads to building bases composed of 14 modes for the solid part, 6 for the fluid part, and 18 for the thermal part. \autoref{fig:modes_multiparametric_Star_Sep} depicts the number of added modes for each computation. We notice that the fluid and thermal bases only increase every 49 computations, which corresponds to a change in the permeability $\permeability$ value. The solid basis requires more modes to describe the solution accurately. Modes are frequently added, but many computations only require an update of solid modes to reach convergence. Moreover, never more than 7 modes are added during a computation, which is sensible regarding the number of modes at the end of the first computation.

It is worth considering whether the solver's performance remains stable when the number of degrees of freedom increases. As explained in \cite{scanff_2021,malleval_2025}, the LATIN-PGD solver resolves a similar number of linear systems as the Newton-Raphson method. Increasing the spatial resolution even leads to slightly fewer linear systems being resolved than with the Newton–Raphson method. However, the number of local stages also increases slightly. Although the analysis was conducted on elasto-visco-plastic problems, it could also be shown for multiphysics problems that the LATIN-PGD approach maintains good scalability with increasing spatial resolution.

\newpage

\section{Conclusion}\label{sec5}

This contribution presents a strategy based on the LATIN-PGD method for solving nonlinear parametrized problems in thermo-poroelasticity. A reduced-order modeling technique such as PGD proves particularly suitable for optimization studies requiring many simulations. Results demonstrate the accuracy of the LATIN-PGD solver in nonlinear thermo-poroelasticity problems on a standard benchmark. The solver also applies to more complex problems and provides consistent solutions. The parametric studies, dealing with material variabilities, reveal substantial computation time savings. Moreover, by constructing each basis independently, the algorithm avoids introducing insignificant modes to describe a given physics. The bases evolve differently for each physics due to the different dynamics at stake in thermo-poroelasticity. \\
Future work thus entails dealing with different discretization for each physics. While \cite{almani_2016} deals with multiple time discretizations in poroelasticity, \cite{dureisseix_2006} presents a multiscale approach based on mortar methods applied to thermo-viscoelasticity. The next stage of future work will combine data-based physics with model-based one. Some multimodel approaches have already been tackled in the frame of multiphysics problems \cite{beckert_2000, bahmani_2021} but do not consider the case where a model is of low fidelity. For such contexts, PGD could prove relevant for enriching one physics basis.

\backmatter

\section*{Declarations}
\begin{itemize}
    \item Availability of data and materials \\
    No data or material was used for the research described in the article.
    \item Competing interests \\
    The authors declare that they have no known competing financial interests or personal relationships that could have appeared to influence the work reported in this paper.
    \item Authors' contributions \\
    E. F.: Writing – original draft, Software, Methodology, Investigation, Conceptualization. P.-A. B.: Review \& editing, Supervision, Methodology, Conceptualization. D. N.: Review \& editing, Supervision, Methodology, Conceptualization. F. L.: Review \& editing, Supervision, Methodology, Conceptualization.
    \item Acknowledgements \\
    The authors gratefully acknowledge Thomas Verbeke, Research Engineer at LMPS, for his support in the development of the ROMlab simulation platform. This work was performed using HPC resources from the ``Mésocentre'' computing center of CentraleSupélec and \'Ecole Normale Supérieure Paris-Saclay supported by CNRS and Région \^Ile-de-France \url{https://mesocentre.universite-paris-saclay.fr/}. The authors gratefully acknowledge the support from the German Research Foundation (DFG) within the International Research Training Group 2657 (IRTG 2657) entitled ``Computational Mechanics Techniques in High Dimensions'' (Grant No. 433082249).
\end{itemize}

\newpage

\begin{appendices}

\section{Notations used in thermo-poroelasticity}\label{secA1}

\begin{table}[ht!]
    \caption{Notations used for a generic thermo-poroelasticity problem}\label{tab:notations}
    \begin{tabular}{c c c}
        \toprule
        \textbf{Symbol} & \textbf{Definition} & \textbf{Units} \\
        \midrule
        & \textit{Fields} & \\
        \midrule
        $\U$ & Displacement & [$\qty{}{\meter}$] \\
        $\Eps$ & Strain tensor & [-] \\
        $\Sig$ & Stress tensor & [$\qty{}{\pascal}$] \\
        $\p$ & Pore pressure & [$\qty{}{\pascal}$] \\
        $\Z$ & Pore pressure gradient & [$\qty{}{\pascal\per\meter}$] \\
        $\vect{V}$ & Darcy's velocity & [$\qty{}{\meter\per\second}$] \\
        $\W$ & Opposite of Darcy's velocity & [$\qty{}{\meter\per\second}$] \\
        $\q$ & Rate of fluid accumulation & [$\qty{}{\per\second}$] \\
        $\T$ & Temperature increment & [$\qty{}{\kelvin}$] \\
        $\X$ & Temperature gradient & [$\qty{}{\kelvin\per\meter}$] \\
        $\qth$ & Heat flux & [$\qty{}{\watt\per\square\meter}$] \\
        $\Y$ & Opposite of the heat flux & [$\qty{}{\watt\per\square\meter}$] \\
        $\R$ & Generalized heat source & [$\qty{}{\watt\per\cubic\meter}$] \\
        \midrule
        &\textit{Material parameters} & \\
        \midrule
        $\Hooke$ & Hooke's operator & [$\qty{}{\pascal}$] \\
        $E$ & Young's modulus & [$\qty{}{\pascal}$] \\
        $\nu$ & Poisson's coefficient & [-] \\
        $\porosity$ & Porosity & [-] \\
        $\drainedbulkmod$ & Drained bulk modulus & [$\qty{}{\pascal}$] \\
        $\bulkmodflu/\bulkmodsol$ & Bulk modulus of the fluid / solid skeleton& [$\qty{}{\pascal}$] \\
        $\biotmod$ & Biot modulus & [$\qty{}{\pascal}$] \\
        $\intpermeability$ & Intrinsic permeability & [$\qty{}{\square\meter}$] \\
        $\fluidvisco$ & Fluid viscosity & [$\qty{}{\pascal\second}$] \\
        $\hydconductivity$ & Hydraulic conductivity & [$\qty{}{\meter\per\second}$] \\
        $\permeability$ & Permeability & [$\qty{}{\cubic\meter\second\per\kilogram}$] \\
        $\couplP$ & Biot coefficient & [-] \\
        $\densityflu / \densitysol$ & Density of the fluid / solid & [$\qty{}{\kilogram\per\cubic\meter}$] \\
        $\heatflu / \heatsol$ & Heat capacity of the fluid / solid & [$\qty{}{\joule\per\kilogram\per\kelvin}$] \\
        $\genheatcapa$ & Generalized heat capacity of the porous medium & [$\qty{}{\joule\per\cubic\meter\per\kelvin}$] \\
        $\thermconductivity$ & Thermal conductivity of the porous medium & [$\qty{}{\watt\per\meter\per\kelvin}$] \\
        $\thermconductivityflu / \thermconductivitysol$ & Thermal conductivity of the fluid / solid & [$\qty{}{\watt\per\meter\per\kelvin}$] \\
        $\expcoefflu / \expcoefsol$ & Thermal expansion coefficient of the fluid / solid & [$\qty{}{\per\kelvin}$] \\
        $\expcoef$ & Thermal expansion coefficient of the porous media & [$\qty{}{\per\kelvin}$] \\
        $\couplT$ & Coupling coefficient between solid and thermal parts & [$\qty{}{\pascal\per\kelvin}$] \\
        \midrule
        & \textit{Initial and boundary conditions} & \\
        \midrule
        $\Ud$ & Displacement prescribed on $\bord{u}$ & [$\qty{}{\meter}$] \\
        $\Fd$ & Loading prescribed on $\bord{F}$ & [$\qty{}{\newton}$] \\
        $\fd$ & Body force prescribed in $\dom$ & [$\qty{}{\newton\per\cubic\meter}$] \\
        $\pd$ & Pore pressure prescribed on $\bord{p}$ & [$\qty{}{\pascal}$] \\
        $\fluxwd$ & Fluid flux prescribed on $\bord{w}$ & [$\qty{}{\meter\per\second}$] \\
        $\po$ & Initial pore pressure & [$\qty{}{\pascal}$] \\
        $\Td$ & Temperature increment prescribed on $\bord{\theta}$ & [$\qty{}{\kelvin}$] \\
        $\qthd$ & Heat flux prescribed on $\bord{q}$ & [$\qty{}{\watt\per\square\meter}$] \\
        $\Rd$ & Heat source prescribed in $\dom$ & [$\qty{}{\watt\per\cubic\meter}$] \\
        $\To$ & Initial temperature increment & [$\qty{}{\kelvin}$]\\
        \botrule
    \end{tabular}
\end{table}

\section{Material parameters for porous ceramics used in the ceramic membrane Star-Sep}\label{secB2}

\begin{table}[ht!]
    \caption{Material parameters for porous ceramics}
    \label{tab:para_mat_star_sep}
    \centering
    \begin{tabular}{c c c}
    \toprule
    Parameters & Value (SI units) & Value (new units) \\
    \hline
    Young modulus $E$ & $\qty{55}{\giga\pascal}$ & $\qty{78.57}{M \: L^{-1} \:S^{-2}}$\\
    Poisson's coefficient $\nu$ & $\qty{0.2}{}$ & $\qty{0.2}{}$\\
    Porosity $\porosity$ & $\qty{0.2}{}$ & $\qty{0.2}{}$\\
    Biot coefficient $\couplP$ & $\qty{0.7}{}$ & $\qty{0.7}{}$\\
    Drained bulk modulus $\drainedbulkmod$ & $\qty{30.6}{\giga\pascal}$ & $\qty{43.7}{M\: L^{-1}\: S^{-2}}$ \\
    Fluid bulk modulus $\bulkmodflu$ & $\qty{2.15}{\giga\pascal}$& $\qty{3.07}{M\: L^{-1}\: S^{-2}}$  \\
    Solid bulk modulus $\bulkmodsol$ & $\qty{102}{\giga\pascal}$ & $\qty{146}{M\: L^{-1}\: S^{-2}}$  \\
    Biot modulus $\biotmod$ & $\qty{10.2}{\giga\pascal}$ &  $\qty{14.6}{M\: L^{-1}\:  S^{-2}}$ \\
    Intrinsic permeability $ \intpermeability$ & $\qty{1.4e-13}{\square\meter}$ & $\qty{1.4e-13}{L^2}$ \\
    Fluid viscosity $\fluidvisco$ & $\qty{1e-3}{\pascal \second}$ & $\qty{1.43e-12}{M\: L^{-1}\: S^{-1}}$\\
    Permeability $\permeability$ & $\qty{1.4e-10}{\cubic\meter\second\per\kilogram}$  & $\qty{9.80e-2}{L^{3}\: S\: M^{-1}}$\\
    Fluid density $\densityflu$ & $\qty{1000}{\kilogram\per\cubic\meter}$ & $\qty{1.43e-6}{M L^{-3}}$\\
    Solid density $\densitysol$ & $\qty{1500}{\kilogram\per\cubic\meter}$ & $\qty{2.14e-6}{M L^{-3}}$\\
    Fluid heat capacity $\heatflu$ & $\qty{4182}{\joule\per\kilogram\per\kelvin}$ & $\qty{1.23e6}{L^2\: S^{-2}\: T^{-1}}$ \\
    Solid heat capacity $\heatsol$ & $\qty{840}{\joule\per\kilogram\per\kelvin}$ & $\qty{2.46e5}{L^2 S^{-2}\: T^{-1}}$\\
    Generalized heat capacity $\genheatcapa$ & $\qty{2.1e6}{\joule\per\cubic\meter\per\kelvin}$ & $\qty{8.79e-1}{M\: L^{-1}\: S^{-2}\: T^{-1}}$ \\
    Fluid thermal conductivity $\thermconductivityflu$ & $\qty{0.61}{\watt\per\meter\per\kelvin}$ & $\qty{2.55e-7}{M\: L S^{-3}\: T^{-1}}$ \\
    Solid thermal conductivity $\thermconductivitysol$ & $\qty{90}{\watt\per\meter\per\kelvin}$ & $\qty{3.77e-5}{M\: L\: S^{-3}\: T^{-1}}$ \\
    Thermal conductivity $\thermconductivity$ & $\qty{72.1}{\watt\per\meter\per\kelvin}$ & $\qty{3.02e-5}{M\: L\: S^{-3}\: T^{-1}}$ \\
    Fluid thermal expansion coefficient $\expcoefflu$ & $\qty{2.6e-4}{\per\kelvin}$ & $\qty{7.62e-2}{T^{-1}}$ \\
    Solid thermal expansion coefficient $\expcoefsol$ & $\qty{8e-6}{\per\kelvin}$ & $\qty{2.34e-3}{T^{-1}}$\\
    Thermal expansion coefficient $\expcoef$ & $\qty{5.84e-5}{\per\kelvin}$ & $\qty{1.71e-2}{T^{-1}}$ \\
    Coupling coefficient $\couplT$ & $\qty{7.34e5}{\pascal\per\kelvin}$ & $\qty{3.07e-1}{M\: L^{-1}\: S^{-2}\: T^{-1}}$\\
    \bottomrule
    \end{tabular}
\end{table}




\end{appendices}

\newpage

\bibliography{sn-bibliography}


\begin{thebibliography}{54}
\ifx \bisbn   \undefined \def \bisbn  #1{ISBN #1}\fi
\ifx \binits  \undefined \def \binits#1{#1}\fi
\ifx \bauthor  \undefined \def \bauthor#1{#1}\fi
\ifx \batitle  \undefined \def \batitle#1{#1}\fi
\ifx \bjtitle  \undefined \def \bjtitle#1{#1}\fi
\ifx \bvolume  \undefined \def \bvolume#1{\textbf{#1}}\fi
\ifx \byear  \undefined \def \byear#1{#1}\fi
\ifx \bissue  \undefined \def \bissue#1{#1}\fi
\ifx \bfpage  \undefined \def \bfpage#1{#1}\fi
\ifx \blpage  \undefined \def \blpage #1{#1}\fi
\ifx \burl  \undefined \def \burl#1{\textsf{#1}}\fi
\ifx \doiurl  \undefined \def \doiurl#1{\url{https://doi.org/#1}}\fi
\ifx \betal  \undefined \def \betal{\textit{et al.}}\fi
\ifx \binstitute  \undefined \def \binstitute#1{#1}\fi
\ifx \binstitutionaled  \undefined \def \binstitutionaled#1{#1}\fi
\ifx \bctitle  \undefined \def \bctitle#1{#1}\fi
\ifx \beditor  \undefined \def \beditor#1{#1}\fi
\ifx \bpublisher  \undefined \def \bpublisher#1{#1}\fi
\ifx \bbtitle  \undefined \def \bbtitle#1{#1}\fi
\ifx \bedition  \undefined \def \bedition#1{#1}\fi
\ifx \bseriesno  \undefined \def \bseriesno#1{#1}\fi
\ifx \blocation  \undefined \def \blocation#1{#1}\fi
\ifx \bsertitle  \undefined \def \bsertitle#1{#1}\fi
\ifx \bsnm \undefined \def \bsnm#1{#1}\fi
\ifx \bsuffix \undefined \def \bsuffix#1{#1}\fi
\ifx \bparticle \undefined \def \bparticle#1{#1}\fi
\ifx \barticle \undefined \def \barticle#1{#1}\fi
\bibcommenthead
\ifx \bconfdate \undefined \def \bconfdate #1{#1}\fi
\ifx \botherref \undefined \def \botherref #1{#1}\fi
\ifx \url \undefined \def \url#1{\textsf{#1}}\fi
\ifx \bchapter \undefined \def \bchapter#1{#1}\fi
\ifx \bbook \undefined \def \bbook#1{#1}\fi
\ifx \bcomment \undefined \def \bcomment#1{#1}\fi
\ifx \oauthor \undefined \def \oauthor#1{#1}\fi
\ifx \citeauthoryear \undefined \def \citeauthoryear#1{#1}\fi
\ifx \endbibitem  \undefined \def \endbibitem {}\fi
\ifx \bconflocation  \undefined \def \bconflocation#1{#1}\fi
\ifx \arxivurl  \undefined \def \arxivurl#1{\textsf{#1}}\fi
\csname PreBibitemsHook\endcsname

\bibitem[\protect\citeauthoryear{Andreozzi et~al.}{2019}]{andreozzi_2019}
\begin{barticle}
\bauthor{\bsnm{Andreozzi}, \binits{A.}},
\bauthor{\bsnm{Iasiello}, \binits{M.}},
\bauthor{\bsnm{Netti}, \binits{P.A.}}:
\batitle{A thermoporoelastic model for fluid transport in tumour tissues}.
\bjtitle{Journal of the Royal Society Interface}
\bvolume{16}(\bissue{154}),
\bfpage{20190030}
(\byear{2019})
\end{barticle}
\endbibitem

\bibitem[\protect\citeauthoryear{Cui et~al.}{2018}]{cui_2018}
\begin{barticle}
\bauthor{\bsnm{Cui}, \binits{W.}},
\bauthor{\bsnm{David M.~Potts}, \binits{D.M.}},
\bauthor{\bsnm{Zdravković}, \binits{L.}},
\bauthor{\bsnm{Gawecka}, \binits{K.A.}},
\bauthor{\bsnm{Taborda}, \binits{D.M.G.}}:
\batitle{An alternative coupled thermo-hydro-mechanical finite element
  formulation for fully saturated soils}.
\bjtitle{Computers and Geotechnics}
\bvolume{94},
\bfpage{22}--\blpage{30}
(\byear{2018})
\doiurl{10.1016/j.compgeo.2017.08.011}
\end{barticle}
\endbibitem

\bibitem[\protect\citeauthoryear{Jin and Peterson}{2025}]{jin_2025}
\begin{barticle}
\bauthor{\bsnm{Jin}, \binits{Z.-H.}},
\bauthor{\bsnm{Peterson}, \binits{M.L.}}:
\batitle{A thermo-poroelasticity model for partially saturated porous media}.
\bjtitle{International Journal of Engineering Science}
\bvolume{208},
\bfpage{104196}
(\byear{2025})
\doiurl{10.1016/j.ijengsci.2024.104196}
\end{barticle}
\endbibitem

\bibitem[\protect\citeauthoryear{Giot et~al.}{2018}]{giot_2018}
\begin{barticle}
\bauthor{\bsnm{Giot}, \binits{R.}},
\bauthor{\bsnm{Granet}, \binits{S.}},
\bauthor{\bsnm{Faivre}, \binits{M.}},
\bauthor{\bsnm{Massoussi}, \binits{N.}},
\bauthor{\bsnm{Huang}, \binits{J.}}:
\batitle{A transversely isotropic thermo-poroelastic model for claystone:
  parameter identification and application to a 3d underground structure}.
\bjtitle{Geomechanics and Geoengineering}
\bvolume{13}(\bissue{4}),
\bfpage{246}--\blpage{263}
(\byear{2018})
\doiurl{10.1080/17486025.2018.1445874}
\end{barticle}
\endbibitem

\bibitem[\protect\citeauthoryear{Wang et~al.}{2021}]{wang_2021}
\begin{barticle}
\bauthor{\bsnm{Wang}, \binits{Y.}},
\bauthor{\bsnm{Zhang}, \binits{Z.}},
\bauthor{\bsnm{Ghassemi}, \binits{A.}}:
\batitle{Modeling of thermo-poroelasticity by using discretized virtual
  internal bond}.
\bjtitle{Geothermics}
\bvolume{91},
\bfpage{102017}
(\byear{2021})
\doiurl{10.1016/j.geothermics.2020.102017}
\end{barticle}
\endbibitem

\bibitem[\protect\citeauthoryear{Ordonez~Egas et~al.}{2023}]{ordonez_2023}
\begin{barticle}
\bauthor{\bsnm{Ordonez~Egas}, \binits{A.C.}},
\bauthor{\bsnm{Tardieu}, \binits{N.}},
\bauthor{\bsnm{Kruse}, \binits{C.}},
\bauthor{\bsnm{Ruiz}, \binits{D.}},
\bauthor{\bsnm{Granet}, \binits{S.}}:
\batitle{Scalable block preconditioners for saturated thermo-hydro-mechanics
  problems}.
\bjtitle{Advanced Modeling and Simulation in Engineering Sciences}
\bvolume{10}(\bissue{1}),
\bfpage{10}
(\byear{2023})
\doiurl{10.1186/s40323-023-00245-z} .
Accessed 2024-07-02
\end{barticle}
\endbibitem

\bibitem[\protect\citeauthoryear{Zhang and Rui}{2024}]{zhang_coupling_2024}
\begin{barticle}
\bauthor{\bsnm{Zhang}, \binits{J.}},
\bauthor{\bsnm{Rui}, \binits{H.}}:
\batitle{A coupling of {Galerkin} and mixed finite element methods for the
  quasi-static thermo-poroelasticity with nonlinear convective transport}.
\bjtitle{Journal of Computational and Applied Mathematics}
\bvolume{441},
\bfpage{115672}
(\byear{2024})
\doiurl{10.1016/j.cam.2023.115672} .
Accessed 2024-04-09
\end{barticle}
\endbibitem

\bibitem[\protect\citeauthoryear{McLean and Espinoza}{2024}]{mclean_2024}
\begin{barticle}
\bauthor{\bsnm{McLean}, \binits{M.L.}},
\bauthor{\bsnm{Espinoza}, \binits{D.N.}}:
\batitle{An open source fem code for solving coupled thermo-poroelastoplastic
  processes}.
\bjtitle{Open Geomechanics}
\bvolume{5}(\bissue{1}),
\bfpage{19}
(\byear{2024})
\end{barticle}
\endbibitem

\bibitem[\protect\citeauthoryear{Felippa and Geers}{1988}]{felippa_1988}
\begin{barticle}
\bauthor{\bsnm{Felippa}, \binits{C.A.}},
\bauthor{\bsnm{Geers}, \binits{T.L.}}:
\batitle{Partitioned analysis for coupled mechanical systems}.
\bjtitle{Engineering Computations}
\bvolume{5}(\bissue{2}),
\bfpage{123}--\blpage{133}
(\byear{1988})
\doiurl{10.1108/eb023730} .
\bcomment{Publisher: MCB UP Ltd}.
Accessed 2024-01-23
\end{barticle}
\endbibitem

\bibitem[\protect\citeauthoryear{Felippa et~al.}{2001}]{felippa_2001}
\begin{barticle}
\bauthor{\bsnm{Felippa}, \binits{C.A.}},
\bauthor{\bsnm{Park}, \binits{K.C.}},
\bauthor{\bsnm{Farhat}, \binits{C.}}:
\batitle{Partitioned analysis of coupled mechanical systems}.
\bjtitle{Computer Methods in Applied Mechanics and Engineering}
\bvolume{190}(\bissue{24}),
\bfpage{3247}--\blpage{3270}
(\byear{2001})
\doiurl{10.1016/S0045-7825(00)00391-1} .
Accessed 2024-01-17
\end{barticle}
\endbibitem

\bibitem[\protect\citeauthoryear{Vila-Chã et~al.}{2023}]{vila-cha_2023}
\begin{barticle}
\bauthor{\bsnm{Vila-Chã}, \binits{J.L.P.}},
\bauthor{\bsnm{Couto~Carneiro}, \binits{A.M.}},
\bauthor{\bsnm{Ferreira}, \binits{B.P.}},
\bauthor{\bsnm{Andrade~Pires}, \binits{F.M.}}:
\batitle{A numerical assessment of partitioned implicit methods for
  thermomechanical problems}.
\bjtitle{Computers \& Structures}
\bvolume{277-278},
\bfpage{106969}
(\byear{2023})
\doiurl{10.1016/j.compstruc.2022.106969} .
Accessed 2024-01-22
\end{barticle}
\endbibitem

\bibitem[\protect\citeauthoryear{Brun et~al.}{2020}]{brun_monolithic_2020}
\begin{barticle}
\bauthor{\bsnm{Brun}, \binits{M.K.}},
\bauthor{\bsnm{Ahmed}, \binits{E.}},
\bauthor{\bsnm{Berre}, \binits{I.}},
\bauthor{\bsnm{Nordbotten}, \binits{J.M.}},
\bauthor{\bsnm{Radu}, \binits{F.A.}}:
\batitle{Monolithic and splitting solution schemes for fully coupled
  quasi-static thermo-poroelasticity with nonlinear convective transport}.
\bjtitle{Computers \& Mathematics with Applications}
\bvolume{80}(\bissue{8}),
\bfpage{1964}--\blpage{1984}
(\byear{2020})
\doiurl{10.1016/j.camwa.2020.08.022} .
Accessed 2024-06-27
\end{barticle}
\endbibitem

\bibitem[\protect\citeauthoryear{Kim}{2018}]{kim_2018}
\begin{barticle}
\bauthor{\bsnm{Kim}, \binits{J.}}:
\batitle{Unconditionally stable sequential schemes for all-way coupled
  thermoporomechanics: Undrained-adiabatic and extended fixed-stress splits}.
\bjtitle{Computer Methods in Applied Mechanics and Engineering}
\bvolume{341},
\bfpage{93}--\blpage{112}
(\byear{2018})
\doiurl{10.1016/j.cma.2018.06.030}
\end{barticle}
\endbibitem

\bibitem[\protect\citeauthoryear{Benner et~al.}{2015}]{benner_2015}
\begin{barticle}
\bauthor{\bsnm{Benner}, \binits{P.}},
\bauthor{\bsnm{Gugercin}, \binits{S.}},
\bauthor{\bsnm{Willcox}, \binits{K.}}:
\batitle{A survey of projection-based model reduction methods for parametric
  dynamical systems}.
\bjtitle{SIAM review}
\bvolume{57}(\bissue{4}),
\bfpage{483}--\blpage{531}
(\byear{2015})
\end{barticle}
\endbibitem

\bibitem[\protect\citeauthoryear{Chatterjee}{2000}]{chatterjee_2000}
\begin{botherref}
\oauthor{\bsnm{Chatterjee}, \binits{A.}}:
An introduction to the proper orthogonal decomposition.
Current science,
808--817
(2000)
\end{botherref}
\endbibitem

\bibitem[\protect\citeauthoryear{Florez}{2017}]{florez_linear_2017}
\begin{bchapter}
\bauthor{\bsnm{Florez}, \binits{H.}}:
\bctitle{Linear {Thermo}-{Poroelasticity} and {Geomechanics}}.
In: \bbtitle{Finite {Element} {Method} - {Simulation}, {Numerical} {Analysis}
  and {Solution} {Techniques}}.
\bpublisher{IntechOpen}, \blocation{???}
(\byear{2017}).
\doiurl{10.5772/intechopen.71873} .
\burl{https://www.intechopen.com/chapters/57902}
Accessed 2024-03-21
\end{bchapter}
\endbibitem

\bibitem[\protect\citeauthoryear{Maday}{2006}]{maday_2006}
\begin{botherref}
\oauthor{\bsnm{Maday}, \binits{Y.}}:
Reduced basis method for the rapid and reliable solution of partial
  differential equations
(2006)
\end{botherref}
\endbibitem

\bibitem[\protect\citeauthoryear{Quarteroni et~al.}{2011}]{quarteroni_2011}
\begin{barticle}
\bauthor{\bsnm{Quarteroni}, \binits{A.}},
\bauthor{\bsnm{Rozza}, \binits{G.}},
\bauthor{\bsnm{Manzoni}, \binits{A.}}:
\batitle{Certified reduced basis approximation for parametrized partial
  differential equations and applications}.
\bjtitle{Journal of Mathematics in Industry}
\bvolume{1},
\bfpage{1}--\blpage{49}
(\byear{2011})
\end{barticle}
\endbibitem

\bibitem[\protect\citeauthoryear{Ryckelynck et~al.}{2006}]{ryckelynck_2006}
\begin{barticle}
\bauthor{\bsnm{Ryckelynck}, \binits{D.}},
\bauthor{\bsnm{Chinesta}, \binits{F.}},
\bauthor{\bsnm{Cueto}, \binits{E.}},
\bauthor{\bsnm{Ammar}, \binits{A.}}:
\batitle{On the a priori model reduction: overview and recent developments}.
\bjtitle{Archives of Computational methods in Engineering}
\bvolume{13},
\bfpage{91}--\blpage{128}
(\byear{2006})
\end{barticle}
\endbibitem

\bibitem[\protect\citeauthoryear{Ladevèze}{1999}]{ladeveze_1999}
\begin{bbook}
\bauthor{\bsnm{Ladevèze}, \binits{P.}}:
\bbtitle{Nonlinear {Computational} {Structural} {Mechanics}: {New} {Approaches}
  and {Non}-{Incremental} {Methods} of {Calculation}}.
\bsertitle{Mechanical {Engineering} {Series}}.
\bpublisher{Springer},
\blocation{New York, NY}
(\byear{1999}).
\doiurl{10.1007/978-1-4612-1432-8} .
\burl{http://link.springer.com/10.1007/978-1-4612-1432-8}
Accessed 2024-01-31
\end{bbook}
\endbibitem

\bibitem[\protect\citeauthoryear{Chinesta et~al.}{2011}]{chinesta_2011}
\begin{barticle}
\bauthor{\bsnm{Chinesta}, \binits{F.}},
\bauthor{\bsnm{Ladevèze}, \binits{P.}},
\bauthor{\bsnm{Cueto}, \binits{E.}}:
\batitle{A {Short} {Review} on {Model} {Order} {Reduction} {Based} on {Proper}
  {Generalized} {Decomposition}}.
\bjtitle{Archives of Computational Methods in Engineering}
\bvolume{18}(\bissue{4}),
\bfpage{395}--\blpage{404}
(\byear{2011})
\doiurl{10.1007/s11831-011-9064-7} .
Accessed 2024-07-25
\end{barticle}
\endbibitem

\bibitem[\protect\citeauthoryear{Beringhier et~al.}{2010}]{beringhier_2010}
\begin{barticle}
\bauthor{\bsnm{Beringhier}, \binits{M.}},
\bauthor{\bsnm{Gueguen}, \binits{M.}},
\bauthor{\bsnm{Grandidier}, \binits{J.C.}}:
\batitle{Solution of strongly coupled multiphysics problems using space-time
  separated representations—application to thermoviscoelasticity}.
\bjtitle{Archives of Computational Methods in Engineering}
\bvolume{17},
\bfpage{393}--\blpage{401}
(\byear{2010})
\doiurl{10.1007/s11831-010-9050-5}
\end{barticle}
\endbibitem

\bibitem[\protect\citeauthoryear{Qin et~al.}{2015}]{qin_2015}
\begin{barticle}
\bauthor{\bsnm{Qin}, \binits{Z.}},
\bauthor{\bsnm{Talleb}, \binits{H.}},
\bauthor{\bsnm{Ren}, \binits{Z.}}:
\batitle{A proper generalized decomposition-based solver for nonlinear
  magnetothermal problems}.
\bjtitle{IEEE Transactions on Magnetics}
\bvolume{52}(\bissue{2}),
\bfpage{1}--\blpage{9}
(\byear{2015})
\end{barticle}
\endbibitem

\bibitem[\protect\citeauthoryear{Schuler et~al.}{2022}]{schuler_2022}
\begin{barticle}
\bauthor{\bsnm{Schuler}, \binits{L.}},
\bauthor{\bsnm{Chamoin}, \binits{L.}},
\bauthor{\bsnm{Khatir}, \binits{Z.}},
\bauthor{\bsnm{Berkani}, \binits{M.}},
\bauthor{\bsnm{Ouhab}, \binits{M.}},
\bauthor{\bsnm{Degrenne}, \binits{N.}}:
\batitle{Iterative pgd model reduction for the strongly-coupled
  thermomechanical analysis of crack propagation in power electronic modules}.
\bjtitle{Computational Mechanics}
\bvolume{70},
\bfpage{407}--\blpage{424}
(\byear{2022})
\doiurl{10.1007/s00466-022-02173-y}
\end{barticle}
\endbibitem

\bibitem[\protect\citeauthoryear{Pruliere et~al.}{2010}]{pruliere_2010}
\begin{barticle}
\bauthor{\bsnm{Pruliere}, \binits{E.}},
\bauthor{\bsnm{Chinesta}, \binits{F.}},
\bauthor{\bsnm{Ammar}, \binits{A.}}:
\batitle{On the deterministic solution of multidimensional parametric models
  using the proper generalized decomposition}.
\bjtitle{Mathematics and Computers in Simulation}
\bvolume{81}(\bissue{4}),
\bfpage{791}--\blpage{810}
(\byear{2010})
\end{barticle}
\endbibitem

\bibitem[\protect\citeauthoryear{Chinesta et~al.}{2010}]{chinesta_2010}
\begin{barticle}
\bauthor{\bsnm{Chinesta}, \binits{F.}},
\bauthor{\bsnm{Ammar}, \binits{A.}},
\bauthor{\bsnm{Cueto}, \binits{E.}}:
\batitle{Recent advances and new challenges in the use of the proper
  generalized decomposition for solving multidimensional models}.
\bjtitle{Archives of Computational methods in Engineering}
\bvolume{17}(\bissue{4}),
\bfpage{327}--\blpage{350}
(\byear{2010})
\end{barticle}
\endbibitem

\bibitem[\protect\citeauthoryear{Chamoin et~al.}{2017}]{chamoin_2017}
\begin{barticle}
\bauthor{\bsnm{Chamoin}, \binits{L.}},
\bauthor{\bsnm{Pled}, \binits{F.}},
\bauthor{\bsnm{Allier}, \binits{P.-E.}},
\bauthor{\bsnm{Ladevèze}, \binits{P.}}:
\batitle{A posteriori error estimation and adaptive strategy for pgd model
  reduction applied to parametrized linear parabolic problems}.
\bjtitle{Computer Methods in Applied Mechanics and Engineering}
\bvolume{327},
\bfpage{118}--\blpage{146}
(\byear{2017})
\end{barticle}
\endbibitem

\bibitem[\protect\citeauthoryear{Ladevèze}{1985}]{ladeveze_1985}
\begin{barticle}
\bauthor{\bsnm{Ladevèze}, \binits{P.}}:
\batitle{Sur une famille d'algorithmes en mécanique des structures}.
\bjtitle{Sur une famille d'algorithmes en mécanique des structures}
\bvolume{300}(\bissue{2}),
\bfpage{41}--\blpage{44}
(\byear{1985}).
\bcomment{Place: Paris Publisher: Gauthier-Villars}
\end{barticle}
\endbibitem

\bibitem[\protect\citeauthoryear{Dureisseix et~al.}{2003}]{dureisseix_2003}
\begin{barticle}
\bauthor{\bsnm{Dureisseix}, \binits{D.}},
\bauthor{\bsnm{Ladevèze}, \binits{P.}},
\bauthor{\bsnm{Schrefler}, \binits{B.A.}}:
\batitle{A {LATIN} computational strategy for multiphysics problems:
  application to poroelasticity}.
\bjtitle{International Journal for Numerical Methods in Engineering}
\bvolume{56}(\bissue{10}),
\bfpage{1489}--\blpage{1510}
(\byear{2003})
\doiurl{10.1002/nme.622} .
\bcomment{\_eprint: https://onlinelibrary.wiley.com/doi/pdf/10.1002/nme.622}.
Accessed 2024-02-23
\end{barticle}
\endbibitem

\bibitem[\protect\citeauthoryear{Wurtzer et~al.}{2024}]{wurtzer_2024}
\begin{barticle}
\bauthor{\bsnm{Wurtzer}, \binits{F.}},
\bauthor{\bsnm{Néron}, \binits{D.}},
\bauthor{\bsnm{Boucard}, \binits{P.-A.}}:
\batitle{A modular model-order reduction approach for the solution of
  parametrized strongly-coupled thermo-mechanical problems}.
\bjtitle{Finite Elements in Analysis and Design}
\bvolume{239},
\bfpage{104207}
(\byear{2024})
\doiurl{10.1016/j.finel.2024.104207} .
Accessed 2024-07-03
\end{barticle}
\endbibitem

\bibitem[\protect\citeauthoryear{Néron and Dureisseix}{2007}]{neron_2007}
\begin{barticle}
\bauthor{\bsnm{Néron}, \binits{D.}},
\bauthor{\bsnm{Dureisseix}, \binits{D.}}:
\batitle{A computational strategy for thermo-poroelastic structures with a
  time-space interface coupling}.
\bjtitle{International Journal for Numerical Methods in Engineering}
\bvolume{75}(\bissue{9}),
\bfpage{1053}--\blpage{1084}
(\byear{2007})
\doiurl{10.1002/nme.2283} .
\bcomment{Publisher: Wiley}.
Accessed 2024-02-23
\end{barticle}
\endbibitem

\bibitem[\protect\citeauthoryear{Boucard and Champaney}{2003}]{boucard_2003}
\begin{barticle}
\bauthor{\bsnm{Boucard}, \binits{P.-A.}},
\bauthor{\bsnm{Champaney}, \binits{L.}}:
\batitle{{A suitable computational strategy for the parametric analysis of
  problems with multiple contact}}.
\bjtitle{{International Journal for Numerical Methods in Engineering}}
\bvolume{57}(\bissue{9}),
\bfpage{1259}--\blpage{1281}
(\byear{2003})
\doiurl{10.1002/nme.724}
\end{barticle}
\endbibitem

\bibitem[\protect\citeauthoryear{Boucard and Ladev{\`e}ze}{1999}]{boucard_1999}
\begin{barticle}
\bauthor{\bsnm{Boucard}, \binits{P.-A.}},
\bauthor{\bsnm{Ladev{\`e}ze}, \binits{P.}}:
\batitle{Application of the latin method to multiple nonlinear structure
  computations (in french)}.
\bjtitle{Revue Européenne des Éléments Finis}
\bvolume{8}(\bissue{8}),
\bfpage{903}--\blpage{920}
(\byear{1999})
\doiurl{10.1080/12506559.1999.10511417}
\end{barticle}
\endbibitem

\bibitem[\protect\citeauthoryear{Heyberger et~al.}{2012}]{heyberger_2012}
\begin{barticle}
\bauthor{\bsnm{Heyberger}, \binits{C.}},
\bauthor{\bsnm{Boucard}, \binits{P.-A.}},
\bauthor{\bsnm{Néron}, \binits{D.}}:
\batitle{Multiparametric analysis within the proper generalized decomposition
  framework}.
\bjtitle{Computational Mechanics}
\bvolume{49}(\bissue{3}),
\bfpage{277}--\blpage{289}
(\byear{2012})
\doiurl{10.1007/s00466-011-0646-x} .
Accessed 2025-01-29
\end{barticle}
\endbibitem

\bibitem[\protect\citeauthoryear{Scanff et~al.}{2022}]{scanff_2022}
\begin{barticle}
\bauthor{\bsnm{Scanff}, \binits{R.}},
\bauthor{\bsnm{Néron}, \binits{D.}},
\bauthor{\bsnm{Ladevèze}, \binits{P.}},
\bauthor{\bsnm{Barabinot}, \binits{P.}},
\bauthor{\bsnm{Cugnon}, \binits{F.}},
\bauthor{\bsnm{Delsemme}, \binits{J.-P.}}:
\batitle{Weakly-invasive latin-pgd for solving time-dependent non-linear
  parametrized problems in solid mechanics}.
\bjtitle{Computer Methods in Applied Mechanics and Engineering}
\bvolume{396},
\bfpage{114999}
(\byear{2022})
\doiurl{10.1016/j.cma.2022.114999}
\end{barticle}
\endbibitem

\bibitem[\protect\citeauthoryear{Daby-Seesaram et~al.}{2025}]{daby_2025}
\begin{barticle}
\bauthor{\bsnm{Daby-Seesaram}, \binits{A.}},
\bauthor{\bsnm{Néron}, \binits{D.}},
\bauthor{\bsnm{Charbonnel}, \binits{P.-E.}},
\bauthor{\bsnm{Fau}, \binits{A.}}:
\batitle{Model-order reduction framework for non-linear dynamics problems
  involving multiple non-parametrised loading configurations for damage
  assessment}.
\bjtitle{Computational Mechanics}
(\byear{2025})
\doiurl{10.1007/s00466-024-02586-x} .
Accessed 2025-01-29
\end{barticle}
\endbibitem

\bibitem[\protect\citeauthoryear{Heyberger et~al.}{2013}]{heyberger_2013}
\begin{barticle}
\bauthor{\bsnm{Heyberger}, \binits{C.}},
\bauthor{\bsnm{Boucard}, \binits{P.-A.}},
\bauthor{\bsnm{N{\'e}ron}, \binits{D.}}:
\batitle{A rational strategy for the resolution of parametrized problems in the
  pgd framework}.
\bjtitle{Computer Methods in Applied Mechanics and Engineering}
\bvolume{259},
\bfpage{40}--\blpage{49}
(\byear{2013})
\end{barticle}
\endbibitem

\bibitem[\protect\citeauthoryear{N{\'e}ron et~al.}{2015}]{neron_2015}
\begin{barticle}
\bauthor{\bsnm{N{\'e}ron}, \binits{D.}},
\bauthor{\bsnm{Boucard}, \binits{P.-A.}},
\bauthor{\bsnm{Relun}, \binits{N.}}:
\batitle{Time-space pgd for the rapid solution of 3d nonlinear parametrized
  problems in the many-query context}.
\bjtitle{International Journal for Numerical Methods in Engineering}
\bvolume{103}(\bissue{4}),
\bfpage{275}--\blpage{292}
(\byear{2015})
\end{barticle}
\endbibitem

\bibitem[\protect\citeauthoryear{Biot}{1941}]{biot_1941}
\begin{barticle}
\bauthor{\bsnm{Biot}, \binits{M.A.}}:
\batitle{General {Theory} of {Three}‐{Dimensional} {Consolidation}}.
\bjtitle{Journal of Applied Physics}
\bvolume{12}(\bissue{2}),
\bfpage{155}--\blpage{164}
(\byear{1941})
\doiurl{10.1063/1.1712886} .
Accessed 2025-02-27
\end{barticle}
\endbibitem

\bibitem[\protect\citeauthoryear{Coussy}{2004}]{coussy_2004}
\begin{bbook}
\bauthor{\bsnm{Coussy}, \binits{O.}}:
\bbtitle{Poromechanics}.
\bpublisher{John Wiley \& Sons}, \blocation{???}
(\byear{2004})
\end{bbook}
\endbibitem

\bibitem[\protect\citeauthoryear{Lewis and
  Schrefler}{1999}]{lewis_schrefler_1999}
\begin{barticle}
\bauthor{\bsnm{Lewis}, \binits{R.W.}},
\bauthor{\bsnm{Schrefler}, \binits{B.A.}}:
\batitle{The {Finite} {Element} {Method} in the {Static} and {Dynamic}
  {Deformation} and {Consolidation} of {Porous} {Media}}.
\bjtitle{Meccanica}
\bvolume{34}(\bissue{3}),
\bfpage{231}--\blpage{232}
(\byear{1999})
\doiurl{10.1023/A:1004546808159} .
Accessed 2024-04-03
\end{barticle}
\endbibitem

\bibitem[\protect\citeauthoryear{Nonino et~al.}{2022}]{nonino_2022}
\begin{barticle}
\bauthor{\bsnm{Nonino}, \binits{M.}},
\bauthor{\bsnm{Ballarin}, \binits{F.}},
\bauthor{\bsnm{Rozza}, \binits{G.}},
\bauthor{\bsnm{Maday}, \binits{Y.}}:
\batitle{Projection {Based} {Semi}-{Implicit} {Partitioned} {Reduced} {Basis}
  {Method} for {Fluid}-{Structure} {Interaction} {Problems}}.
\bjtitle{Journal of Scientific Computing}
\bvolume{94}(\bissue{1}),
\bfpage{4}
(\byear{2022})
\doiurl{10.1007/s10915-022-02049-6} .
Accessed 2024-03-20
\end{barticle}
\endbibitem

\bibitem[\protect\citeauthoryear{Scanff et~al.}{2021}]{scanff_2021}
\begin{barticle}
\bauthor{\bsnm{Scanff}, \binits{R.}},
\bauthor{\bsnm{Nachar}, \binits{S.}},
\bauthor{\bsnm{Boucard}, \binits{P.-A.}},
\bauthor{\bsnm{Néron}, \binits{D.}}:
\batitle{A {Study} on the {LATIN}-{PGD} {Method}: {Analysis} of {Some}
  {Variants} in the {Light} of the {Latest} {Developments}}.
\bjtitle{Archives of Computational Methods in Engineering}
\bvolume{28}(\bissue{5}),
\bfpage{3457}--\blpage{3473}
(\byear{2021})
\doiurl{10.1007/s11831-020-09514-1} .
Accessed 2024-02-07
\end{barticle}
\endbibitem

\bibitem[\protect\citeauthoryear{Néron and Dureisseix}{2008}]{neron_2008}
\begin{barticle}
\bauthor{\bsnm{Néron}, \binits{D.}},
\bauthor{\bsnm{Dureisseix}, \binits{D.}}:
\batitle{A computational strategy for poroelastic problems with a time
  interface between coupled physics}.
\bjtitle{International Journal for Numerical Methods in Engineering}
\bvolume{73}(\bissue{6}),
\bfpage{783}--\blpage{804}
(\byear{2008})
\doiurl{10.1002/nme.2091} .
\bcomment{Publisher: Wiley}.
Accessed 2024-02-23
\end{barticle}
\endbibitem

\bibitem[\protect\citeauthoryear{Aboustit et~al.}{1985}]{aboustit_1985}
\begin{barticle}
\bauthor{\bsnm{Aboustit}, \binits{B.L.}},
\bauthor{\bsnm{Advani}, \binits{S.H.}},
\bauthor{\bsnm{Lee}, \binits{J.K.}}:
\batitle{Variational principles and finite element simulations for
  thermo-elastic consolidation}.
\bjtitle{International Journal for Numerical and Analytical Methods in
  Geomechanics}
\bvolume{9}(\bissue{1}),
\bfpage{49}--\blpage{69}
(\byear{1985})
\doiurl{10.1002/nag.1610090105} .
\bcomment{\_eprint:
  https://onlinelibrary.wiley.com/doi/pdf/10.1002/nag.1610090105}.
Accessed 2024-06-26
\end{barticle}
\endbibitem

\bibitem[\protect\citeauthoryear{Pogacnik et~al.}{2011}]{pogacnik_2011}
\begin{bchapter}
\bauthor{\bsnm{Pogacnik}, \binits{J.}},
\bauthor{\bsnm{Leary}, \binits{P.}},
\bauthor{\bsnm{Malin}, \binits{P.}}:
\bctitle{Implementation and numerical analysis of fullycoupled non-isothermal
  fluid flow through a deformable porous medium}.
In: \bbtitle{Proceedings NZ Geothermal Workshop, Auckland},
pp. \bfpage{21}--\blpage{23}
(\byear{2011})
\end{bchapter}
\endbibitem

\bibitem[\protect\citeauthoryear{Sanavia et~al.}{2008}]{sanavia_2008}
\begin{barticle}
\bauthor{\bsnm{Sanavia}, \binits{L.}},
\bauthor{\bsnm{Fran{\c{c}}ois}, \binits{B.}},
\bauthor{\bsnm{Bortolotto}, \binits{R.}},
\bauthor{\bsnm{Luison}, \binits{L.}},
\bauthor{\bsnm{Laloui}, \binits{L.}}, \betal:
\batitle{Finite element modelling of thermo-elasto-plastic water saturated
  porous materials}.
\bjtitle{Journal of Theoretical and Applied Mechanics}
\bvolume{38},
\bfpage{7}--\blpage{24}
(\byear{2008})
\end{barticle}
\endbibitem

\bibitem[\protect\citeauthoryear{Sandhu et~al.}{1977}]{sandhu_1977}
\begin{barticle}
\bauthor{\bsnm{Sandhu}, \binits{R.S.}},
\bauthor{\bsnm{Liu}, \binits{H.}},
\bauthor{\bsnm{Singh}, \binits{K.J.}}:
\batitle{Numerical performance of some finite element schemes for analysis of
  seepage in porous elastic media}.
\bjtitle{International Journal for Numerical and Analytical Methods in
  Geomechanics}
\bvolume{1}(\bissue{2}),
\bfpage{177}--\blpage{194}
(\byear{1977})
\doiurl{10.1002/nag.1610010205} .
\bcomment{\_eprint:
  https://onlinelibrary.wiley.com/doi/pdf/10.1002/nag.1610010205}.
Accessed 2024-03-22
\end{barticle}
\endbibitem

\bibitem[\protect\citeauthoryear{Aboustit et~al.}{1982}]{aboustit_1982}
\begin{bchapter}
\bauthor{\bsnm{Aboustit}, \binits{B.L.}},
\bauthor{\bsnm{Advani}, \binits{S.H.}},
\bauthor{\bsnm{Lee}, \binits{J.K.}},
\bauthor{\bsnm{Sandhu}, \binits{R.S.}}:
\bctitle{Finite element evaluations of thermo-elastic consolidation}.
In: \bbtitle{ARMA US Rock Mechanics/geomechanics Symposium},
p. \bfpage{82}
(\byear{1982}).
\bcomment{ARMA}
\end{bchapter}
\endbibitem

\bibitem[\protect\citeauthoryear{Malleval et~al.}{2025}]{malleval_2025}
\begin{barticle}
\bauthor{\bsnm{Malleval}, \binits{P.-E.}},
\bauthor{\bsnm{Scanff}, \binits{R.}},
\bauthor{\bsnm{Néron}, \binits{D.}}:
\batitle{Advancing industrial finite element software: Developing model order
  reduction for nonlinear transient thermal problems}.
\bjtitle{Finite Elements in Analysis and Design}
\bvolume{244},
\bfpage{104299}
(\byear{2025})
\doiurl{10.1016/j.finel.2024.104299}
\end{barticle}
\endbibitem

\bibitem[\protect\citeauthoryear{Almani et~al.}{2016}]{almani_2016}
\begin{barticle}
\bauthor{\bsnm{Almani}, \binits{T.}},
\bauthor{\bsnm{Kumar}, \binits{K.}},
\bauthor{\bsnm{Dogru}, \binits{A.}},
\bauthor{\bsnm{Singh}, \binits{G.}},
\bauthor{\bsnm{Wheeler}, \binits{M.F.}}:
\batitle{Convergence analysis of multirate fixed-stress split iterative schemes
  for coupling flow with geomechanics}.
\bjtitle{Computer Methods in Applied Mechanics and Engineering}
\bvolume{311},
\bfpage{180}--\blpage{207}
(\byear{2016})
\end{barticle}
\endbibitem

\bibitem[\protect\citeauthoryear{Dureisseix and
  Bavestrello}{2006}]{dureisseix_2006}
\begin{barticle}
\bauthor{\bsnm{Dureisseix}, \binits{D.}},
\bauthor{\bsnm{Bavestrello}, \binits{H.}}:
\batitle{Information transfer between incompatible finite element meshes:
  application to coupled thermo-viscoelasticity}.
\bjtitle{Computer Methods in Applied Mechanics and Engineering}
\bvolume{195}(\bissue{44-47}),
\bfpage{6523}--\blpage{6541}
(\byear{2006})
\end{barticle}
\endbibitem

\bibitem[\protect\citeauthoryear{Beckert}{2000}]{beckert_2000}
\begin{barticle}
\bauthor{\bsnm{Beckert}, \binits{A.}}:
\batitle{Coupling fluid (cfd) and structural (fe) models using finite
  interpolation elements}.
\bjtitle{Aerospace Science and technology}
\bvolume{4}(\bissue{1}),
\bfpage{13}--\blpage{22}
(\byear{2000})
\end{barticle}
\endbibitem

\bibitem[\protect\citeauthoryear{Bahmani and Sun}{2021}]{bahmani_2021}
\begin{barticle}
\bauthor{\bsnm{Bahmani}, \binits{B.}},
\bauthor{\bsnm{Sun}, \binits{W.C.}}:
\batitle{A kd-tree-accelerated hybrid data-driven/model-based approach for
  poroelasticity problems with multi-fidelity multi-physics data}.
\bjtitle{Computer Methods in Applied Mechanics and Engineering}
\bvolume{382},
\bfpage{113868}
(\byear{2021})
\end{barticle}
\endbibitem

\end{thebibliography}

\end{document}